\documentclass[10pt]{article}
\usepackage{amsmath,amssymb,tikz}
\usepackage{graphicx}
 \allowdisplaybreaks
\def\be{\begin{equation}}
\def\ee{\end{equation}}
\def\ba#1\ea{\begin{align}#1\end{align}}
\def\bg#1\eg{\begin{gather}#1\end{gather}}
\def\bm#1\em{\begin{multline}#1\end{multline}}
\def\bmd#1\emd{\begin{multlined}#1\end{multlined}}

\def\a{\alpha}
\def\b{\beta}

\def\d{\delta}

\def\l{\lambda}

\def\m{\mu}
\def\n{\nu}

\def\r{\rho}

\def\t{\tau}

\def\nn{\nonumber}

\def\({\left(}
\def\){\right)}
\def\[{\left[}
\def\]{\right]}

\def \xfu {{\overset{\text{\tiny{(1)}}}{X}{}^{i}}}

\def \xfd {{\overset{\text{\tiny{(1)}}}{X}{}_i}}
\def \xsd {{\overset{\text{\tiny{(2)}}}{X}{}_i}}
\def \gaterm {{\overset{\text{\tiny{(0)}}}{\Gamma}{}^{m}_{ki}\ \overset{\text{\tiny{(1)}}}{X}{}^i\overset{\text{\tiny{(1)}}}{X}{}^k\overset{\text{\tiny{(1)}}}{X}{}_m}}
\def \be {\begin{equation}}
\def \ee {\end{equation}}
\def \ba {\begin{array}}
\def \ea {\end{array}}
\def \bea{\begin{eqnarray}}
\def \eea{\end{eqnarray}}
\def \nn {\nonumber}

\def \a {\alpha}
\def \b {\beta}

\def \d {\delta}

\def \m {\mu}
\def \n {\nu}

\def \l {\lambda}

\def \r {\rho}

\def \t {\tau}

\def\bea{\begin{eqnarray}}
\def\eea{\end{eqnarray}}
\newcommand{\eq}[1]{(\ref{#1})}
\newcommand{\bit}{\begin{itemize}}  \newcommand{\eit}{\end{itemize}}
\newcommand{\ben}{\begin{enumerate}}  \newcommand{\een}{\end{enumerate}}

\long\def\symbolfootnote[#1]#2{\begingroup%
\def\thefootnote{\fnsymbol{footnote}}\footnote[#1]{#2}\endgroup}

\newcommand{\nthu}{{\it Department of Physics, National Tsing-Hua
  University,
Hsinchu 30013, Taiwan}}

\newcommand{\ncts}{{\it Physics Division, National Center for Theoretical Sciences,
National Tsing-Hua University, Hsinchu 30013, Taiwan}}

\begin{document}
\thispagestyle{empty}
\hfill{ NCTS-TH/1702} 
\begin{center}

~\vspace{20pt}

{\Large\bf On New Proposal for Holographic BCFT}

\vspace{25pt}

Chong-Sun Chu${}^{1,2}$\symbolfootnote[1]{Email:~\sf cschu@phys.nthu.edu.tw}
Rong-Xin Miao ${}^1$\symbolfootnote[3]{Email:~\sf
  miaorongxin.physics@gmail.com} and
Wu-Zhong Guo${}^2$ \symbolfootnote[2]{Email:~\sf wzguo@cts.nthu.edu.tw}

\vspace{10pt}${}^1$\nthu

\vspace{10pt}${}^2$\ncts\\

\vspace{2cm}

\begin{abstract}
This paper is an extended version of our short letter on a new proposal for
holographic boundary conformal field, i.e.,  BCFT.  By using
the Penrose-Brown-Henneaux (PBH) transformation, we
successfully obtain the expected boundary Weyl anomaly.
The obtained boundary central charges satisfy naturally a c-like
theorem holographically. We then develop an approach of holographic
renormalization
for BCFT, and reproduce the correct boundary Weyl anomaly.
This provides a non-trivial check of our proposal. We also investigate the
holographic entanglement entropy of BCFT and find
that our proposal
gives the expected orthogonal condition
that the minimal surface must be normal to the spacetime boundaries if they
intersect. This is
another support for our proposal.
We also find that
the entanglement entropy depends on the boundary conditions of BCFT and
the distance to the boundary; and that the entanglement wedge behaves a phase
transition, which is important for the self-consistency of
AdS/BCFT. Finally, we show that the proposal of arXiv:1105.5165 is too
restrictive that it always make vanishing some of the boundary
central charges.

\end{abstract}

\end{center}

\newpage
\setcounter{footnote}{0}
\setcounter{page}{1}

\tableofcontents

\section{Introduction}

BCFT is a conformal field theory defined on a manifold $M$  with a
boundary $P$ with suitable boundary conditions. It has important
applications in string theory and  condensed matter physics near
boundary
critical behavior \cite{Cardy:2004hm}. In the spirit of AdS/CFT
\cite{Maldacena:1997re},  Takayanagi \cite{Takayanagi:2011zk}
proposes to extend the $d$ dimensional manifold $M$ to a $d+1$
dimensional asymptotically AdS space $N$ so that $\partial N= M\cup
Q$, where $Q$ is a $d$ dimensional manifold which satisfies $\partial
Q=\partial M=P$. 
We mention that the presence of the boundary $Q$ is very natural 
from the point of view of
the UV/IR relation  \cite{uvir} of AdS/CFT correspondence since the presence of
boundary in  the field theory introduce an IR cutoff and 
this can be naturally implemented in the bulk with the presence of a
boundary.
Conformal invariance on $M$ requires that $N$ is part of AdS space.
The key point of holographic BCFT is 
thus to determine the
location of boundary $Q$ in the bulk.
For interesting developments
of BCFT and related topics please see,
for example, \cite{Nozaki:2012qd,Fujita:2011fp,
  Jensen:2015swa,Erdmenger:2015spo,Erdmenger:2014xya,Miyaji:2014mca,
  Jensen:2013lxa,Estes:2014hka,Gaiotto:2014gha,
  Fursaev:2016inw,Berthiere:2016ott,He:2014gva}.

The gravitational action for holographic BCFT is given by
\cite{Takayanagi:2011zk,Nozaki:2012qd}  (taking $16\pi G_N =1$)
\begin{eqnarray}\label{action1}
I=\int_N \sqrt{G} (R-2 \Lambda) + 2\int_Q \sqrt{h} (K-T)+ 2\int_M
\sqrt{g}\ K+2\int_P \sqrt{\sigma}\ \theta
\end{eqnarray}
where $T$ is a constant and $\theta=\arccos(n_M.n_Q)$ is the
supplementary angle between the boundaries $M$ and $Q$, which makes a
well-defined variational principle on the corner $P$
\cite{Hayward:1993my}.  Notice that $T$  can be regarded as the
holographic dual of boundary conditions of BCFT
since it affects the boundary
entropy 
(and also the boundary central charges, see  (\ref{3dBWA1},\ref{4dBWA1}) below)
which are closely related to the
boundary conditions of BCFT \cite{Takayanagi:2011zk,Nozaki:2012qd}. 
Considering the variation of the on-shell action, we have
\begin{eqnarray}\label{daction1}
\delta I&=&-\int_Q \sqrt{h}
\left(K^{\alpha\beta}-(K-T)h^{\alpha\beta}\right) \delta
h_{\alpha\beta}
-\int_M \sqrt{g} (K^{ij}-K h^{ij})\delta g_{ij}+\int_P
\sqrt{\sigma}\ \theta \sigma^{ab}\delta \sigma_{ab}.
\end{eqnarray}
For conformal boundary conditions in CFT, 
Takayanagi \cite{Takayanagi:2011zk} proposes to impose Dirichlet  boundary
condition on 
$M$ and $P$, $\delta g_{ij}|_M=\delta \sigma_{ab}|_P=0$, but Neumann
boundary condition on $Q$. And the position of the boundary $Q$ is
determined by the Neumann boundary condition
\begin{eqnarray}\label{NBC}
K^{\alpha\beta}-(K-T)h^{\alpha\beta}=0.
\end{eqnarray}
For more general boundary conditions which break boundary conformal
invariance even locally,
\cite{Takayanagi:2011zk,Nozaki:2012qd} propose
to add matter fields on $Q$ and replace eq.(\ref{NBC}) by
\begin{eqnarray}\label{NBC1}
K_{\alpha\beta}-K h_{\alpha\beta}=\frac{1}{2} T^Q_{\alpha\beta},
\end{eqnarray}
where we have included $2 T h_{\alpha\beta} $ in the matter stress
tensor $ T^Q_{\alpha\beta}$ on
$Q$. 
For geometrical shape of $M$ with high symmetry such
as the case of a disk or half plane, 
(\ref{NBC}) fixes the location of $Q$ and produces many elegant
results for  BCFT \cite{Takayanagi:2011zk,Nozaki:2012qd,Fujita:2011fp}. 
However 
since 
$Q$ is of co-dimension one and its shape is determined by a single 
embedding function,
(\ref{NBC}) gives too many constraints and
in general
 there is no solution in a given spacetime such as AdS. 
 On the other hand, of course,
 one expect to have
 well-defined BCFT with general boundaries.

To solve this problem,
\cite{Nozaki:2012qd} propose to take into
account backreactions of $Q$. For 3d BCFT, they show that one can
indeed find perturbative solution to (\ref{NBC}) if one take into
account backreactions to the bulk spacetime. In other words although
not all the shapes
of boundary $P$ are allowed by (\ref{NBC}) in a given spacetime, 
by carefully tuning the spacetime (which is a solution to Einstein equations)
one can always make (\ref{NBC}) consistent for
any given shape.
However, it is still a little restrictive since 
one has to change both the ambient spacetime and the position of $Q$
for different boundaries of the BCFT.


As motivated in \cite{Takayanagi:2011zk,Nozaki:2012qd},
the conditions \eq{NBC} and \eq{NBC1}  
are natural from the point of view of braneworld
scenario, and so is the backreaction. However from a practical point
of view,  it is
not entirely satisfactory since one
has a large freedom to choose the matter fields as long as they
satisfy various energy conditions. As a result, it seems one can put the
boundary $Q$ at almost any position as one likes. 
Besides,
it is unappealing that
the holographic dual depends on the details of matters
on another boundary $Q$.
Finally, although eq.(\ref{NBC1}) could have solutions 
for general shapes
by
tuning the matters, it is 
actually too strong since
as we will prove in the appendix it 
always makes vanishing some of the central charges 
in the boundary Weyl anomaly.  In a recent work \cite{Miao:2017gyt}, we
propose a new holographic
dual of BCFT with $Q$ determined by a new condition \eq{mixedBCN2}.
This condition is consistent and provides a unified treatment to 
general shapes 
of $P$. 
Besides, as we will show below, it yields the expected boundary
contributions to the Weyl anomaly.

 Instead of imposing
Neumann boundary condition (\ref{NBC}), we suggest to impose the mixed
boundary conditions on $Q$  \cite{Miao:2017gyt}:
\begin{eqnarray}\label{mixedBCN}
&&(K^{\alpha\beta}-(K-T)h^{\alpha\beta})\Pi_{+\alpha\beta}^{\ \ \  \alpha'\beta'}=0,\\
&&\Pi_{-\alpha\beta}^{\ \ \ \alpha'\beta'}\delta h_{\alpha'\beta'}=0.\label{mixedBCD}
\end{eqnarray}
where $\Pi_{+\alpha\beta}^{\ \ \ \alpha'\beta'}$ and
$\Pi_{-\alpha\beta}^{\ \ \ \alpha'\beta'}$ 
are the projection operators satisfying
$\Pi_{+\alpha\beta}^{\ \ \ \ \alpha'\beta'}+\Pi_{-\alpha\beta}^{\ \ \ \ \alpha'\beta'}
=\delta_{\alpha}^{\alpha'}\delta_{\beta}^{\beta'}$
and $\Pi_{\pm \alpha\beta}^{\ \ \ \  \alpha'\beta'}\Pi_{\pm
  \alpha'\beta'}^{\ \ \  \ \ \alpha_1\beta_1}=\Pi_{\pm
  ij}^{\ \ \  \ \alpha_1\beta_1}$. Since we could impose at most one
condition to fix the location of the co-dimension one surface $Q$, we
require $\Pi_{+\alpha\beta}^{\ \ \  \alpha'\beta'}= A_{\alpha\beta}B^{
  \alpha'\beta'}$ and $tr AB=1$ from $\Pi_{+}\Pi_{+}=\Pi_{+}$. Now the
mixed boundary condition (\ref{mixedBCN}) becomes
\begin{eqnarray}\label{mixedBCN1}
(K^{\alpha\beta}-(K-T)h^{\alpha\beta})A_{\alpha\beta}=0,
\end{eqnarray}
where $A_{\alpha\beta}$ are non-zero tensors to be determined. It is natural to require that
eq. (\ref{mixedBCN1}) to be linear in $K$ so that it is a second order
differential equation for the embedding. 
Thus we propose the choice
$A_{\alpha\beta}=h_{\alpha\beta}$ in \cite{Miao:2017gyt}.  In this paper, we will provide more evidences
for this proposal. Besides, we find that 
the other choices such as
\be \label{otherA}
A_{\alpha\beta}=\lambda_1 h_{\alpha\beta}+ \lambda_2
K_{\alpha\beta}+\lambda_3 R_{\alpha\beta}+ \cdots,
\;\;\;
\lambda_1, \lambda_2 \neq 0, \;\;\; 
\ee 
all lead problems.

To sum up, we propose to use the traceless condition 
\begin{eqnarray}\label{mixedBCN2}
T_{BY}{}^{\alpha}_{\ \alpha}=2(1-d)K+2dT=0
\end{eqnarray}
to determine the boundary $Q$. Here
$T_{\rm BY}{}_{\alpha\beta}=2K_{\alpha\beta}-2(K-T)h_{\alpha\beta}$ 
is the Brown-York stress tensor on $Q$.
In general, it could also depend on the intrinsic curvatures which we
will treat in sect.4. 
A few remarks on \eq{mixedBCN2} are in order.
{\it 1.} It is worth noting that the junction 
condition for a thin shell with spacetime on both sides 
is 
also given by (\ref{NBC1}) \cite{Hayward:1993my}. However, here
$Q$ is the
boundary of spacetime and not a thin shell,
so there is no need to consider
the junction condition. 
{\it 2.} For the same reason,  it is expected that $Q$ has
no back-reaction on the geometry just as the boundary $M$. 
{\it 3.} Eq. (\ref{mixedBCN2}) implies that $Q$ is a
constant mean curvature surface,
which is also of great interests in both
mathematics and physics \cite{Rafael} just as 
the minimal surface. 
{\it 4.} \eq{mixedBCN2}  reduces to the proposal by
\cite{Takayanagi:2011zk} for a disk and half-plane. And it can
reproduce all the results in
\cite{Takayanagi:2011zk,Nozaki:2012qd,Fujita:2011fp}. 
{\it 5.} 
Eq. \eq{mixedBCN2} is a purely geometric equation and
has solutions for arbitrary shapes of boundaries and arbitrary bulk
metrics.
{\it 6.} Very importantly,
our proposal 
gives non-trivial boundary Weyl anomaly, which solves the difficulty met in
\cite{Takayanagi:2011zk,Nozaki:2012qd}.
In fact as we will show in the appendix
the proposal (\ref{NBC1}) of \cite{Takayanagi:2011zk} 
is too restrictive
and always yields $c_2=b_1=0$ for the central charges 
in (\ref{3dBWA},\ref{4dBWA}). Since $b_1$ is expected to satisfy a c-like theorem and
describes the degree of freedom on the boundary, thus it is important
for $b_1$ to be non-zero.

Let us recall that in the presence of boundary,  Weyl anomaly 
of CFT generally pick up a boundary contribution
$\left<T^a_a\right>_P$ in addition to the usual bulk term
$\left<T^i_i\right>_M$, i.e.
$ \left< T^i_i \right>=\left<T^i_i\right>_M
+\delta(x_{\perp})\left<T^a_a\right>_P$,
where $\delta(x_{\perp})$ is a delta function with support on the
boundary $P$. Our proposal yields
the expected boundary Weyl anomaly for 3d and 4d BCFT 
\cite{Herzog:2015ioa,Fursaev:2015wpa,Solodukhin:2015eca}:
\begin{eqnarray}\label{3dBWA}
&&\left<T^a_a\right>_P=c_1 \mathcal{R}+ c_2 \text{Tr}
  \bar{k}^2, \qquad \qquad\qquad \qquad\quad \;\;d=3,\\
&&\left<T^a_a\right>_P=\frac{a}{16\pi^2} E_4^{\rm bdy}+ b_1 \text{Tr}
  \bar{k}^3 + b_2 C^{ac}_{\ \ \ b c}
  \bar{k}_{\ a}^b,\quad d=4,\label{4dBWA}
\end{eqnarray}
where $c_1, c_2, b_1,b_2$ are boundary central charges, $a=2\pi^2$ is
the bulk central charge for 4d CFTs dual to Einstein gravity,  $
\mathcal{R}$ is intrinsic curvature, $\bar{k}_{ab}$ is the traceless
part of extrinsic curvature, $C_{ijkl}$ is the Weyl tensor on $M$ and
$(- E_4^{\rm bdy})$ is the boundary terms of Euler density $E_4$ used to
preserve the topological invariance
\begin{eqnarray}\label{E4bry}
 E_4^{\rm bdy}=4 \left ( 2 \text{Tr}
 (k\mathcal{R})-k\mathcal{R}+\frac{2}{3} \text{Tr} k^3- k \text{Tr}
 k^2+\frac{1}{3}k^3 \right).
\end{eqnarray}
Since $Q$ is not a minimal surface in our case, our results (\ref{3dBWA1},\ref{4dBWA1}) are non-trivial generalizations of the Graham-Witten anomaly \cite{Graham:1999pm} for the submanifold.


The paper is organized as follows. In sect.2, we study PBH transformations in the
presence of submanifold which is not orthogonal to the AdS boundary $M$ and
derive the boundary contributions to holographic Weyl anomaly for 3d and 4d
BCFT. In sect. 3, we investigate the holographic renormalization
for BCFT,
and reproduce the correct boundary Weyl anomaly obtained in sect.2,
which provides a non-trivial check of our proposal. In sect. 4, we consider
the general boundary conditions of BCFT by adding intrinsic
curvature terms on the bulk boundary $Q$.  In sect.5,
we study the holographic entanglement entropy and boundary effects
on entanglement. In sect.6, we discuss the  phase
transition of entanglement wedge, which is important for the self-consistency of
AdS/BCFT. Conclusions and discussions are found in sect. 7.
The paper is finished with three appendices. In appendix A, we give an
independent derivation of the leading and subleading terms
of the embedding function by solving directly our proposed boundary condition
for $Q$. The result agrees with that obtained in sect.2 using the PBH
transformations. In appendix B, we show that the proposal of
\cite{Takayanagi:2011zk}  always make
vanish the central charges $c_2$ and $b_1$ in the boundary Weyl anomaly for 3d
and 4d BCFT.
 In appendix C, we give the details of calculations for the boundary
 contributions to Weyl anomaly.

Notations: $G_{\mu\nu},\  g_{ij},\  h_{\alpha\beta}$ and $\sigma_{ab}$
are the metrics in $N, \ M, \ Q$ and $P$, respectively. We have
$\mu=(1,...,d+1)$, $i=(1,...,d)$, $\alpha=(1,...,d)$ and $a=(1,...,
d-1)$. The curvatures are defined by
$R^{\rho}_{\ \sigma\mu\nu}=\partial_{\mu}\Gamma^{\rho}_{\sigma\nu}+
\Gamma^{\rho}_{\mu\lambda}\Gamma^{\lambda}_{\sigma\nu}-(\mu\leftrightarrow
\nu)$, $R_{\mu\nu}=R^{\rho}_{\ \mu\rho\nu}$ and
$R=R^{\mu}_{\ \mu}$. The
extrinsic curvature on $Q$ are defined by 
$K= \nabla_{\mu}n^{\mu}$, where $n_{\mu}$ is the unit vector normal to
$Q$ and pointing outward from $N$ to $Q$.

Note added: Two weeks after \cite{Miao:2017gyt}, there appears a paper
\cite{Astaneh:2017ghi} which claims that our calculations of boundary
Weyl anomaly (\ref{3dBWA1},\ref{4dBWA1}) are not correct. We find they
have ignored important contributions from the bulk action $I_N$ for 3d
BCFT and the boundary action $I_Q$ for 4d BCFT. After communication
with us, they realize
the problems and reproduce our results  (\ref{3dBWA1},\ref{4dBWA1}) in
a new revision of \cite{Astaneh:2017ghi}. For the convenience of the
reader, we give the details of
our calculations in appendix C.
We also
emphasis here that, from our analysis, it is natural 
to keep $T$ as a free parameter rather than to
set it zero. Otherwise, the corresponding 2d BCFT becomes trivial
since the boundary entropy \cite{Takayanagi:2011zk} is zero when
$T=0$. Besides, we emphasis that, as we previously demonstrated
in section 4, by
allowing intrinsic curvatures terms
on $Q$, one can always make the holographic boundary Weyl anomaly
matches the predictions of 
BCFT with general boundary
conditions.
This may or may not match with the result of free BCFT 
since so far it is not clear whether and
how non-renormalization theorems hold. However in the special case it
holds, e.g. in the presence of supersymmetry,
it just means the parameters of the intrinsic curvature terms are
fixed, which is completely natural due to the presence of more symmetry.


\section{Holographic Boundary Weyl Anomaly}

According to \cite{Schwimmer:2008yh}, the embedding function of the
boundary $Q$ is highly constrained by the asymptotic symmetry of AdS,
and it can be determined by PBH transformations up to some conformal
tensors. By using PBH transformations, we find the leading and
subleading terms of the embedding function for $Q$ are universal and
can be used to derive the boundary contributions to
the Weyl anomaly for
3d and 4d BCFT. It is worth noting that we do not make any assumption
about the location of $Q$ in this approach. So the holographic
derivations of  boundary Weyl anomaly in this section is very strong.

\subsection{PBH transformation}

Let us firstly briefly review PBH transformation in the presence of a
submanifold \cite{Schwimmer:2008yh}. Consider a  $(p+1)$-dimensional
submanifold $\Sigma$ embedded into the $(d+1)$-dimensional bulk $N$
such that it ends on a $p$-dimensional submanifold $\partial \Sigma$
on the $d$-dimensional boundary $M$. Denote the bulk coordinates by
$X^{\mu}=(x^i, \rho)$ and the coordinates on $\Sigma$ by
$\tau^{\alpha}=(y^a,\tau)$ with $i=1,...,d$ and $a=1,...,p$. The
embedding function is given by $X^{\mu}=X^{\mu}(\tau^{\alpha})$.

We consider the bulk metric in the 
FG gauge
\begin{eqnarray}\label{FGmetric}
ds^2=\frac{d\rho^2}{4\rho^2}+\frac{g_{ij}dx^idx^j}{\rho}.
\end{eqnarray}
Here $\rho =0$ denote the boundary of the metric.
It is known that 
if one assume the
metric $g_{ij}$ admits a series expansion in powers of $\rho$,
$g_{ij}=\overset{(0)}{g}_{ij}+\rho \overset{(1)}{g}_{ij}+ \cdots$, 
then $\overset{(1)}{g}_{ij}$ can be fixed by the PBH transformation
\cite{Imbimbo:1999bj}
\footnote{Note that in our notation, the sign of
  curvatures differs from the one of
  \cite{Schwimmer:2008yh,Imbimbo:1999bj} by a minus sign.}
\begin{eqnarray}\label{AdSmetric1}
\overset{(1)}{g}_{ij}=-\frac{1}{d-2}\left(
R^{(0)}_{ij}-\frac{R^{(0)}}{2(d-1)}g^{(0)}_{ij}\right).
\end{eqnarray}
PBH transformations are a special subgroup of diffeomorphism which
preserve the FG gauge:
\begin{eqnarray}\label{bulkPBH1}
&&\delta \rho= -2\rho \sigma(x), \\
&&\delta x^i=a^i=\frac{1}{2}\int_{0}^{\rho}d\rho'
  g^{ij}(x,\rho')\partial_j 
\sigma(x)+a^i_0(x). \label{bulkPBH2}
\end{eqnarray}
Here $\sigma(x)$ is the parameter of Weyl rescalings of the boundary
metric, i.e., $\delta_{\sigma}g^{(0)}_{ij}=2\sigma g^{(0)}_{ij}$ and
$a^i_0(x)$ is the diffeomorphism of the boundary $M$. To keep the
position of $\partial \Sigma$ on $M$, we require that
$a^i_0(x)|_{\partial\Sigma}=0$.

Next let us include the submanifold. 
The metric on $\Sigma$ is given by
\begin{eqnarray}\label{Qmetric}
&&h_{\tau\tau}=\frac{1}{4\tau^2}+\frac{1}{\tau}\partial_{\tau}X^i\partial_{\tau}X^j
  g_{ij}(X,\tau),\\
&&h_{ab}=\frac{1}{\tau}\partial_{a}X^i\partial_{b}X^jg_{ij}(X,\tau).
\end{eqnarray}
To fix the reparametrization invariance on $\Sigma$, 
we chose similarly the gauge fixing condition
\begin{eqnarray}\label{STgauge}
\tau=\rho, \ \  h_{a\tau}=0
\end{eqnarray}
Now under a bulk PBH transformation (\ref{bulkPBH1},\ref{bulkPBH2}),
one needs to make a compensating diffeomorphism on $\Sigma$ 
\cite{Schwimmer:2008yh}  
such that $\delta \rho=\delta \tau$ and 
$\delta h_{a\tau}=\partial_a \tilde{\xi}^{\tau}
h_{\tau\tau}+\partial_{\tau} \tilde{\xi}^{b} h_{ab}=0$
in order to stay in the gauge (\ref{STgauge}). This gives
\begin{eqnarray}\label{killingQ}
\tilde{\xi}^{\tau}=-2\tau \sigma(x) \quad \mbox{and}\quad
\tilde{\xi}^{a}=2\int_{0}^{\tau}d\tau' \tau' h_{\tau\tau} h^{ab}\partial_b \sigma.
\end{eqnarray}
As a result,  $X^i$ changes under PBH
transformation as
\begin{eqnarray}\label{PBHXi}
\delta X^i=\tilde{\xi}^{\alpha}\partial_{\alpha} X^i-a^i,
\end{eqnarray}
where $\tilde{\xi}^\a$ is given by \eq{killingQ} and
 $a^i$ is given by eq.(\ref{bulkPBH2}). 
As in the case of the metric, if one expand the embedding function 
in powers of $\tau$, 
\begin{eqnarray}\label{Xi0}
X^i(\tau, y^a)=\overset{(0)}{X}{}^i(y^a)+\tau
\overset{(2)}{X}{}^i(y^a)+ \cdots,
\end{eqnarray}
the first leading nontrivial term can be fixed by its transformation
properties \cite{Schwimmer:2008yh}. In fact, since
\begin{eqnarray}\label{PBHXi1}
&&\delta\overset{(0)}{X}{}^i=0,\nonumber\\
&&\delta \overset{(2)}{X}{}^i=-2\sigma
  \overset{(2)}{X}{}^i+\frac{1}{2}\overset{(0)}{h}{}^{ab}\partial_a
  \overset{(0)}{X}{}^i\partial_b
  \sigma-\frac{1}{2}\overset{(0)}{g}{}^{ij}\partial_j \sigma,
\end{eqnarray}
one can solve the second equation of \eq{PBHXi1} by
\begin{eqnarray}\label{X1}
\overset{(2)}{X}{}^i=\frac{1}{2p}k^i,
\end{eqnarray}
where $k^i$ is the trace of the extrinsic curvature of $\partial \Sigma$
\begin{eqnarray}\label{ki}
k^i=\overset{(0)}{h}{}^{ab}k^i_{ab}=\overset{(0)}{h}{}^{ab}\left(
\partial_a\partial_b
\overset{(0)}{X}{}^i-\overset{(0)}{\gamma}{}^c_{ab}\partial_c
X^{(0)}{}^i+\overset{(0)}{\Gamma}{}^i_{jk}\partial_a
\overset{(0)}{X}{}^j \partial_b \overset{(0)}{X}{}^k \right),
\end{eqnarray}
$\overset{(0)}{h}{}^{ab}$ is the inverse of $\overset{(0)}{h}{}_{ab}$
which appears in the expansion:
\be
h_{ab} = \frac{1}{\tau}\partial_{a}\overset{(0)}{X}{}^i
\partial_{b}\overset{(0)}{X}{}^jg_{ij}(\overset{(0)}{X},\tau) + \cdots :=
\frac{1}{\tau} \overset{(0)}{h}{}_{ab} + \cdots
\ee
and
$\overset{(0)}{\gamma}{}_{ab}^c$ is the Christoffel symbol for the
induced metric $\overset{(0)}{h}{}_{ab}$.

Now let us focus on our problem with $p= d-1$, $\Sigma=Q$ and
$\partial \Sigma=P$. Inspired by \cite{Takayanagi:2011zk}, we relax
the assumption of \cite{Schwimmer:2008yh} and expand $X^i$ in powers
of $\sqrt{\tau}$ in the presence of a boundary: 
\begin{eqnarray}\label{Xi}
X^i(\tau, y^a)=\overset{(0)}{X}{}^i(y^a)+\sqrt{\tau}
\overset{(1)}{X}{}^i(y^a)+\tau\overset{(2)}{X}{}^i(y^a)+ \cdots
\end{eqnarray}
This means that $\Sigma=Q$ is not orthogonal to the AdS boundary $M$
generally due to the non-zero $\overset{(1)}{X}{}^i(y^a)$. 
Then we have
\begin{eqnarray}\label{hat}
h_{\tau a}&=&\frac{1}{\tau}\partial_{\tau}X^i\partial_{a}X^jg_{ij}(X,\tau)\nonumber\\
&=&\frac{1}{2\tau^{\frac{3}{2}}}\overset{(1)}{X}{}^i
\partial_{a}\overset{(0)}{X}{}^jg^{(0)}_{ij}
+\frac{1}{2\tau}\left(  2
\overset{(2)}{X}{}^i\partial_{a}
\overset{(0)}{X}{}^j\overset{(0)}{g}_{ij}+\overset{(1)}{X}{}^i\partial_{a}
\overset{(1)}{X}{}^j\overset{(0)}{g}_{ij}+
\overset{(1)}{X}{}^i\partial_{a}\overset{(0)}{X}{}^j
\overset{(1)}{X}{}^{k}\partial_k\overset{(0)}{g}_{ij}\right)
+ \cdots. \;\;\;\;
\end{eqnarray}
Imposing the gauge (\ref{STgauge}), we get
\begin{eqnarray}\label{hat1}
&&\overset{(1)}{X}{}^i= |\overset{(1)}{X}|\  n^i,\\
&&h^i_j \overset{(2)}{X}{}^j =-\frac{1}{4}h^{ij}\partial_j
  |\overset{(1)}{X}|^2-\frac{1}{2}h^i_j \overset{(0)}{\Gamma}{}^j_{kl}
  n^k n^l |\overset{(1)}{X}|^2\label{hat2},
\end{eqnarray}
where $n^i$ is the normal vector pointing inside from $P$ to $M$,
$|\overset{(1)}{X}|=\sqrt{\overset{(1)}{X}{}^i\overset{(1)}{X}{}_i}$,
$h_{ij} :=\overset{(0)}{g}_{ij}-n_in_j$ is the zeroth order induced
metric on $\Sigma$, 
$k^i=- n^i k$ and $k=\nabla_i n^i$. It is worth noting that
$\overset{(2)}{X}{}^i$ is on longer a vector due to the appearance of
the affine term in eq.(\ref{hat2}). This is not surprising since we have
imposed the gauge (\ref{STgauge}) which fixes all the
reparametrization of $Q$ except the one acting on $\partial Q=P$
\cite{Schwimmer:2008yh}. One can easily check that $
\overset{(0)}{\Gamma}{}^j_{kl} n^k n^l$ is indeed
covariant under
the residual gauge transformations of the reparametrization of
$P$. Besides, note that coordinates are not vector generally, so
there is no need to require $\overset{(2)}{X}{}^i$ to be a
vector. What must be covariant are the finial results such Weyl
anomaly and entanglement entropy.

Now let us study the transformations of  $X^i$ under PBH. 
From eq.(\ref{PBHXi}), we obtain
\begin{eqnarray}
&&\delta\overset{(0)}{X}{}^i=0,\label{PBHXimine0}\\
&&\delta\overset{(1)}{X}{}^i=-\sigma \overset{(1)}{X}{}^i,\label{PBHXimine1}\\
&&\delta \overset{(2)}{X}{}^i=-2\sigma
  \overset{(2)}{X}{}^i+\frac{1}{2}|\overset{(1)}{X}|^2h^{ij}\partial_j
  \sigma-\frac{1}{2}(1+2|\overset{(1)}{X}|^2)n^in^j\partial_j\sigma.
\label{PBHXimine2}
\end{eqnarray}
Using the following formulas
\begin{eqnarray}\label{affinen}
&& \delta_{\sigma} n^i=-\sigma n^i,\\
&& \delta_{\sigma} k^i=-2\sigma k^i-p\  n^in^j\partial_j \sigma,\label{affinek}\\
&&\delta_{\sigma} \overset{(0)}{\Gamma}{}^i_{jk}=\delta^i_j \nabla_k
  \sigma+\delta^i_k \nabla_j
  \sigma-\overset{(0)}{g}_{jk}\nabla^i\sigma,\label{affine}
\end{eqnarray}
one can easily check that eqs.(\ref{hat1},\ref{hat2}) indeed obey the
transformations (\ref{PBHXimine1},\ref{PBHXimine2}). 
One may also solve \eq{PBHXimine2} directly and obtain for the normal
components of $\overset{(2)}{X}{}^i$ as:
\begin{eqnarray}\label{hat3}
&&n^in_j\overset{(2)}{X}{}^j=
\frac{1+|\overset{(1)}{X}|^2}{2p} k^i-\frac{1}{2}|\overset{(1)}{X}|^2
\overset{(0)}{\Gamma}{}^n_{nn} n^i+c_1 (\frac{k^i}{p}+
\overset{(0)}{\Gamma}{}^n_{nn} n^i).
\end{eqnarray}
Here
$\overset{(0)}{\Gamma}{}^n_{nn}=\overset{(0)}{\Gamma}{}^i_{jk}n_in^jn^k$ 
and
$c_1$ is a parameter to be determined.
Note that
a term proportional to $n^in^j\partial_j
|\overset{(1)}{X}|^2$ from \eq{hat2} drops out automatically in 
\eq{hat3} since 
$|\overset{(1)}{X}(y^a)|$ is functions of only the transverse
coordinates $y^a$, such term vanishes due to the normal derivatives.

As we have mentioned, 
$\overset{(2)}{X}{}^i$ is no longer a vector in the normal sense due
to the gauge fixing (\ref{STgauge}).
Instead, $\overset{(2)}{X}{}^i$ admit some kinds of
deformed covariance
under the remaining diffeomorphism after fixing the FG
gauge (\ref{FGmetric}) in $N$ and world-volume gauge (\ref{STgauge})
on $Q$. It is clear that the remaining diffeomorphism are the ones on
$M$ and $P$. The key point is that,  for every diffeomorphism on $M$,
there exists compensating reparametrization on $Q$ in order to stay
in the  gauge (\ref{STgauge}). As a result, $\overset{(2)}{X}{}^i$ is
covariant
in a certain sense under the combined diffeomorphisms on
$M$ and $Q$. As we will illustrate below,  the deformed gauge symmetry
is useful and it fixes the value of the parameter $\l_1$ to be zero.

Without loss of generality, we consider the Gauss normal coordinates
$X^{i}=(x,y^a)$ on $M$
\begin{eqnarray}\label{BCFTmetric1}
ds_M^2=\overset{(0)}{g}_{ij}dx^idx^j=dx^2+ \left(\sigma_{ab}+2x
k_{ij}+ x^2 q_{ij}+ \cdots \right)dy^a dy^b,
\end{eqnarray}
where $P$ is located at $x=0$, and $Q$ is determined by
\begin{eqnarray}\label{gaussQ}
x=a_1(y) \sqrt{\tau} +a_2(y) \tau+ \cdots
\end{eqnarray}
To satisfy the gauge (\ref{STgauge}), we should choose the coordinates 
on $Q$ carefully. For example, the natural one $\tau^{\alpha}=(y^a,
\tau)$ 
does not work. Instead, we should choose $\tau^{\alpha}=(y'^a, \tau)$
with the embedding functions given by
\begin{eqnarray}\label{gaussembedding1}
&& \rho=\tau, \\
&& x=a_1(y') \sqrt{\tau} +a_2(y') \tau+ \cdots \label{gaussembedding2}\\
&& y^a=y'^a -\frac{1}{4} \sigma^{ab}\partial_b a_1^2(y') \tau+ \cdots
\label{gaussembedding3}
\end{eqnarray}
Notice that $n^i=(1,0, \cdots ,0)$ and $\Gamma^n_{nn}=0$ for 
the Gauss normal coordinates (\ref{BCFTmetric1}). Recall also that
$k=-n_i k^i$, we obtain from  eq.(\ref{hat3})
\begin{eqnarray}\label{gaussa2}
a_2(y')=-\frac{1+a_1^2(y')}{2p} k-c_1 \frac{k}{p}.
\end{eqnarray}
Now let us use the remaining diffeomorphism to fix the parameter
$c_1$. Consider a remaining diffeomorphism
\begin{eqnarray}\label{remainingdiff1}
x=x'+ c x'^2+O(x'^3)
\end{eqnarray}
which keeps the position of $P$ and the gauge
eqs.(\ref{FGmetric},\ref{STgauge}). 
From eqs.(\ref{gaussembedding2},\ref{gaussa2},\ref{remainingdiff1}), we have
\begin{eqnarray}\label{remainingdiff2}
x'=x- c x^2+O(x^3)= a_1(y') \sqrt{\tau}- \left(\frac{1+a_1^2(y')}{2p} 
k + c_1 \frac{k}{p} + c \ a_1^2(y') \right) \tau+ \cdots
\end{eqnarray}
Since the new coordinate $x'$ satisfies the gauge
(\ref{FGmetric},\ref{STgauge}), it must take the form (\ref{hat3})
because of PBH transformations. Substituting $n'^i=(1,0,..,0)$ and
$\Gamma^{n'}_{n'n'}=2c$ into  eq.(\ref{hat3}), we get
\begin{eqnarray}\label{gaussa2b}
a'_2(y')=-\frac{1+a_1^2(y')}{2p} k-c_1 \frac{k}{p} - c
\ a_1^2(y') +2 c c_1
\end{eqnarray}
for the new coordinate $x'$.
Comparing eq.(\ref{gaussa2b}) with the coefficients of $\tau$ in
eq.(\ref{remainingdiff2}), we find that they match if and only if
$c_1=0$. Hence our claim.

As a summary, by using the PBH transformations and 
the covariance under remaining diffeomorphism, we find the leading and
subleading terms of embedding functions are universal and take the
following form
\begin{eqnarray}\label{finhat1}
&&\overset{(1)}{X}{}^i= |\overset{(1)}{X}|\  n^i,\\
&&\overset{(2)}{X}{}^i =\frac{1+|\overset{(1)}{X}|^2}{2p}
  k^i-\frac{1}{4}h^{ij}\partial_j
  |X^{(1)}|^2-\frac{1}{2}\overset{(0)}{\Gamma}{}^i_{nn}|\overset{(1)}{X}|^2
\label{finhat2}
\end{eqnarray}
In the Gauss normal coordinates (\ref{BCFTmetric1}), the embedding
function has very elegant expression
\begin{eqnarray}\label{finGaussx}
x=a_1(y) \sqrt{\tau}-\frac{1+a^2_1(y) }{2p} k\ \tau+ \cdots
\end{eqnarray}
These are the main results of this section. One may still doubt
eq.(\ref{finhat2}) due to the non-covariance. Actually, we can derive
it from the covariant equation (\ref{mixedBCN2}) together with the
gauge (\ref{STgauge}). So it must be covariant under the remaining
diffeomorphism. This is a non-trivial check of our results. Please see
the appendix for the details. Besides, we have checked other choices
of boundary conditions such as eq.(\ref{mixedBCN1}) with
$A_{\alpha\beta}= h_{\alpha\beta}+ \lambda_2 K_{\alpha\beta}+\lambda_3
R_{\alpha\beta}$. They all yield the same results
eqs.(\ref{finhat1},\ref{finhat2},\ref{finGaussx}). This is a strong
support for the universality.

\subsection{Boundary Weyl anomaly}

In this section, we
apply the method of \cite{Henningson:1998gx} to derive the
Weyl anomaly (including the boundary contributions to Weyl anomaly
\cite{Nozaki:2012qd}) as the logarithmic divergent term of the
gravitational action. For our purpose, we focus only on the boundary
Weyl anomaly on $P$ below.

Let us quickly recall our main setup.
Consider the asymptotically AdS metric
\begin{eqnarray}\label{AdSmetric}
ds^2=\frac{dz^2+g_{ij}dx^idx^j}{z^2}
\end{eqnarray}
where $z=\sqrt{\rho}$, $g_{ij}=g^{(0)}_{ij}+z^2 g^{(1)}_{ij}+\cdots$, 
$g^{(0)}_{ij}$ is the metric of BCFT on $M$ and $g^{(1)}_{ij}$,
fixed uniquely by the PBH transformation,  is given by \eq{AdSmetric1}. 
Without loss of generality, we choose Gauss normal coordinates 
for the metric on $M$
\begin{eqnarray}\label{BCFTmetric}
ds_M^2=g^{(0)}_{ij}dx^idx^j=dx^2+ \left(\sigma_{ab}+2x k_{ij}+ x^2
q_{ab}+x^3 l_{ab}+ \cdots \right)dy^a dy^b,
\end{eqnarray}
where the boundary $P$ is located at $x=0$ . 
The bulk boundary $Q$ is given by $x=X(z,y)$. Expanding it in $z$, we have
\begin{eqnarray}\label{Qsuface}
x=a_1 z+ a_2 z^2+\cdots+ ( b_{d+1} \ln z+ a_{d+1}) z^{d+1}+ \cdots
\end{eqnarray}
where $a_i$ and $b_{d+1}$ are functions of $y^a$. By using the 
PBH transformation, we know that $a_2$ is universal and can be
expressed 
in terms of $a_1$ and the extrinsic curvature $k$ through
eq.(\ref{finGaussx}). $a_1$ can be determined by the boundary
condition on $Q$. Noting that
$K^{\alpha}_{\beta}=\frac{a_1}{\sqrt{1+a_1^2}}\delta^{\alpha}_{\beta}+O(z)$,
we get the leading term of  eq.(\ref{mixedBCN1}) as
\begin{eqnarray}\label{leadingBC}
\left(K^{\alpha}_{\beta}-(K-T)\delta^{\alpha}_{\beta}\right)A^{\beta}_{\alpha}=
\left( (1-d)\frac{a_1}{\sqrt{1+a_1^2}}+T \right) A^{\alpha}_{\alpha}+ \cdots=0,
\end{eqnarray}
where $\cdots$ denotes higher order terms in $z$. It is remarkable that
we can solve $a_1$ from eq.(\ref{leadingBC}) without any assumption of
$A_{ij}$ except its trace is nonzero. In other words, we can solve
$a_1$ from the universal part of the boundary conditions. From
eqs.(\ref{finGaussx},\ref{leadingBC}), we finally obtain
\begin{eqnarray}\label{uniTa}
&&T=(d-1)\tanh \rho_*,\quad a_1= \sinh \rho_*,
  \quad a_2=-\frac{\text{Tr}k}{2(d-1)}\cosh^2\rho_*, 
\cdots, 
\end{eqnarray}
where we have re-parameterized the constant $T$ in terms of $\rho_*$,
which can be regarded as the holographic dual of boundary conditions
for BCFT. That is because, as will be clear soon, $\rho_*$  affects
the boundary central charges as the boundary conditions do. It should
be mentioned that one can also obtain $a_1, a_2$ by directly solving
the boundary condition eq.(\ref{mixedBCN2}) or eq.(\ref{mixedBCN1})
with $A_{\alpha\beta}= h_{\alpha\beta}+ \lambda_2
K_{\alpha\beta}+\lambda_3 R_{\alpha\beta}+ \cdots$. They yield the same
results for $(T, a_1, a_2)$ but different results for $(a_3, a_4, \cdots)$.

Now we are ready to derive the boundary Weyl anomaly. For simplicity,
we focus on the case of 3d BCFT and 4d BCFT.  Substituting
eqs.(\ref{AdSmetric}-\ref{uniTa}) into the action (\ref{action1}) and
selecting the logarithmic divergent terms after the integral along $x$
and $z$, we can obtain the boundary Weyl anomaly.  
We note that 
$I_M$ and $I_P$ do not contribute to the logarithmic divergent term
in the action since they have at most singularities in powers of $z^{-1}$  
but there is
no integration alone $z$, thus there is no way for them 
to produce $\log z$ terms. 
We also note that only $a_2$ appears in the final results. The terms
including $a_3$ and $a_4$ automatically cancel each other out. This is
also the case for the holographic Weyl anomaly and universal terms of
entanglement entropy for 4d and 6d CFTs \cite{Miao:2013nfa,Miao:2015iba}. 
After some calculations, we obtain
the boundary Weyl anomaly for 3d and 4d BCFT as
\begin{eqnarray}\label{3dBWA1}
&&\left<T^a_a\right>_P= \sinh \rho_*\  \mathcal{R}-\sinh \rho_*\  \text{Tr}
  \bar{k}^2,\ \ \ \  \ \ \ \ \ \ \  \ \ \ \ \ \ \ \ \ \ \ \ \ \ \ \ \ \ \ \ \ \ \ \ \text{ for 3d
    BCFT},\\
&&\left<T^a_a\right>_P= \frac{1}{8} E_4^{\rm bdy}+ (\cosh(2\rho_*)-\frac{1}{3})
  \text{Tr} \bar{k}^3 -\cosh(2\rho_*) C^{ac}_{\ \ \ b c}
  \bar{k}_{\ a}^b,\ \text{ for 4d BCFT.}\label{4dBWA1}
\end{eqnarray}
which takes the expected conformal invariant form
\cite{Herzog:2015ioa,Fursaev:2015wpa,Solodukhin:2015eca}.   It is
remarkable that the coefficient of $ E_4^{\rm bdy}$ takes the correct
value to  preserve the topological invariance of $E_4$. This is a
non-trivial check of our results. Besides, the 
boundary charges $c_1, b_1$ in (\ref{3dBWA}, \ref{4dBWA})
are expected to satisfy a c-like theorem \cite{Nozaki:2012qd,Jensen:2015swa,Huang:2016rol}. 
As was shown in \cite{Takayanagi:2011zk,Fujita:2011fp}, null energy condition on
$Q$ implies $\rho$ decreases along RG flow. It is also true for us. As a
result, 
eqs.(\ref{3dBWA1}, \ref{4dBWA1}) indeed obey the c-theorem for
boundary charges.
This is also
a support for our results.  Most importantly, our confidence is based
on the above universal derivations, i.e., we do not make any
assumption except the universal part of the boundary conditions on
$Q$. 
Last but not least, we notice that our results (\ref{3dBWA1},\ref{4dBWA1}) are non-trivial generalizations of the Graham-Witten anomaly \cite{Graham:1999pm} for the submanifold, i.e., we find there exists conformal invariant boundary Weyl anomaly for non-minimal surfaces.

We remark that 
based on the results of free CFTs \cite{Fursaev:2015wpa} and the
variational principle, it 
has been suggested that the coefficient of $Ck$
in (\ref{4dBWA1}) is universal for all 4d BCFTs
\cite{Solodukhin:2015eca}. Here we provide 
evidence, based on holography, against this suggestion: 
our results agree with the suggestion of
\cite{Solodukhin:2015eca} for the trivial case $\rho=0$, 
while disagree generally.  As argued in \cite{Huang:2016rol}, the
proposal of \cite{Solodukhin:2015eca} is suspicious. It
means that there could be no independent boundary central charge
related to the Weyl invariant
$\sqrt{\sigma} C^{ac}_{\ \ \ b c} \bar{k}_{\ a}^b$. However, in general, every
Weyl invariant should correspond to an independent central charge,
such as the case for 2d, 4d and 6d CFTs.
Besides, we notice
that the law obeyed by free CFTs usually does not apply to strongly
coupled CFTs. See
\cite{Dong:2016wcf,Lee:2014zaa,Hung:2014npa,Chu:2016tps} for 
examples. 

To summarize, by using the universal term in the embedding functions
eq.(\ref{finGaussx}) and the universal part of the boundary condition
eq.(\ref{mixedBCN1}), we succeed to derive the boundary contributions
to Weyl anomaly for 3d and 4d BCFTs. Since we do not need to assume
the exact position of $Q$, the holographic derivations of boundary
Weyl anomaly here is very strong. On the other hand, since the terms
including $a_3$ and $a_4$ automatically cancel each other out in the
above calculations, so far we cannot distinguish our proposal
(\ref{mixedBCN2}) from the other
possibilities
such as
eq.(\ref{mixedBCN1}) with $A_{\alpha\beta}= h_{\alpha\beta}+ \lambda_2
K_{\alpha\beta}+\lambda_3 R_{\alpha\beta}$. We will solve this
problem in the next section.

\section{Holographic Renormalization of BCFT}

In this section, we develop the holographic renormalization for
BCFT. We find that one should add new kinds of counterterms on
boundary $P$ in order to get finite action. Using this scheme, we
reproduce the correct boundary Weyl anomaly
eqs.(\ref{3dBWA1},\ref{4dBWA1}), which provides a strong support for
our proposal eq.(\ref{mixedBCN2}).

\subsection{3d BCFT}

Let us use the regularized stress tensor \cite{Balasubramanian:1999re}
to study the boundary Weyl anomaly. This method
requires the knowledge
of $(a_3, a_4,\cdots)$ and thus can help us to distinguish the proposal
(\ref{mixedBCN2}) from the other choices. we will focus on the case of
3d BCFT in this subsection.\\
The first step is to find a finite action by adding suitable covariant
counterterms\cite{Balasubramanian:1999re}. We obtain
\begin{eqnarray}\label{regularizedaction1}
I_{\rm ren}&=&\int_N dx^4\sqrt{G} (R-2 \Lambda) + 2\int_Q dx^3\sqrt{h}(
K-T)
+ 2\int_M dx^3\sqrt{g}(K-2-\frac{1}{2}R_M )\nonumber\\
&&+2\int_P dy^2 \sqrt{\sigma}(\theta -\theta_0-K_M),
\end{eqnarray}
where $I_M$ includes the usual counterterms in holographic
renormalization \cite{Balasubramanian:1999re,deHaro:2000vlm},
$\theta_0=\theta(z=0)$ is a constant \cite{Nozaki:2012qd}, $K_M$ is
the Gibbons-Hawking-York term for $R_M$ on $M$. Notice that there is
no freedom to add other counterterms, except some finite terms which
are irrelevant to Weyl anomaly. For example, we may add
$\sqrt{\sigma}R_P$ and $\sqrt{\sigma}K_M^2$ to $I_P$. However, these
terms are invariant under constant Weyl transformations. Thus they do
not contribute to the boundary Weyl Anomaly. In conclusion, the
regularized action (\ref{regularizedaction1}) is unique up to some
irrelevant finite counterterms.

 From the renormalized action, it is straightly to derive the
 Brown-York stress tensor on $P$
\begin{eqnarray}\label{regularizedstresstensor1}
B_{ab}=2(K_{Mab}-K_M \sigma_{ab})+2(\theta -\theta_0)\sigma_{ab}
\end{eqnarray}
In sprint of \cite{Nozaki:2012qd,Balasubramanian:1999re,deHaro:2000vlm}, 
the boundary Weyl anomaly is given by
\begin{eqnarray}\label{HBWAstress1}
\left<T^a_a\right>_P=\lim_{z\to 0} \frac{B^a_a}{z^2}=\lim_{z\to 0}
\frac{4(\theta-\theta_0)-2K_M}{z^2},
\end{eqnarray}
where $\theta=\arccos \frac{x'}{\sqrt{g^{xx}+x'^2}}+O(z^3)$, 
$\theta_0= \arccos(\tanh \rho)$ and $K_M=z\frac{\partial_x
  (\sqrt{g}\sqrt{g^{xx}})}{\sqrt{g}}+O(z^3)$.
Actually since we are interested only in boundary Weyl anomaly, we do
not need to calculate all the components of Brown-York stress tensors
on $P$. Instead, we can play a trick. From the constant Weyl
transformations $\sigma_{ab}\to e^{2\epsilon}\sigma_{ab}$,
$\sqrt{\sigma}\to e^{2\epsilon} \sqrt{\sigma}$, $\theta \to \theta$
and $K_M\to e^{-\epsilon} K_M$, we can read off the  boundary Weyl
normally as
\begin{eqnarray}\label{HBWAstress2}
\int_P dy^2\sqrt{\sigma_0}\left<T^a_a\right>_P=\int_P
dy^2\sqrt{\sigma} \left( 4(\theta-\theta_0)-2K_M \right),
\end{eqnarray}
which agrees with eq.(\ref{HBWAstress1}) exactly.

Substituting eqs.(\ref{AdSmetric}-\ref{uniTa}) into
eq.(\ref{HBWAstress1}), we obtain
\begin{eqnarray}\label{HWAmstress1}
 \left<T^a_a\right>_P=-\frac{1}{4} \text{sech}^2(\rho ) [48 a_3+\sinh
   (\rho ) \left(2 \mathcal{R}+6 q-3 k^2-6 \text{Tr}k^2\right)+\sinh
   (3 \rho ) \left(2 q-k^2-4 \text{Tr}k^2\right)]
\end{eqnarray}
 Comparing eq.(\ref{HWAmstress1}) with eq.(\ref{3dBWA1}), we find that
 they match if and only if
\begin{eqnarray}\label{a3}
a_3=\frac{1}{48} \sinh (\rho ) \left(\cosh (2 \rho ) (-2 \mathcal{R}-4
q+k^2+10 \text{Tr}k^2)-4\mathcal{R}-8 q+3k^2+12 \text{Tr}k^2\right),
\end{eqnarray}
which is exactly the solution to our proposed boundary condition
(\ref{mixedBCN2}). One can check that
eq.(\ref{mixedBCN1}) with the other choices
$A_{\alpha\beta}= h_{\alpha\beta}+ \lambda_2
K_{\alpha\beta}+\lambda_3 R_{\alpha\beta}$ gives different $a_3$ and
thus can be
ruled out. Following the same approach, we can also derive
boundary Weyl anomaly for 4d BCFT, which agrees with eq.(\ref{4dBWA1})
if and only if $a_3$ and $a_4$ are given by the solutions to condition
(\ref{mixedBCN2}). This is a very strong support to the boundary
condition (\ref{mixedBCN2}) we proposed.

To end this section, let us talk more about the stress tensors on
$P$. In general, since the  Brown-York stress tensor on $Q$ is
non-vanishing, we have
\begin{eqnarray}\label{dIre3d}
\delta I_{\rm ren}=\frac{1}{2}\int_M \sqrt{g_0} T_M^{ij} \delta
g^{(0)}_{ij}+ \frac{1}{2}\int_P \sqrt{\sigma_0} T^{ab} \delta
\sigma_{0 ab}+ \frac{1}{2}\int_Q \sqrt{h} T_Q^{\alpha\beta} \delta
h_{\alpha\beta}
\end{eqnarray}
From the viewpoint of BCFT, the variations of effective action should
takes the form
\begin{eqnarray}\label{dIre3dBCFT}
\delta I_{\rm eff}=\frac{1}{2}\int_M \sqrt{g_0} T_M^{ij} \delta
g^{(0)}_{ij}+ \frac{1}{2}\int_P \sqrt{\sigma_0}\left( T_e^{ab} \delta
\sigma_{0 ab}+ J \delta O \right)
\end{eqnarray}
where $J$ and $O$ are the currents and operators on $P$,
respectively. After the integration along $z$ on $Q$, we can identify
$I_{\rm ren}$ with $I_{\rm eff}$. Since $\sigma_{0 ab}=\lim_{z\to
  0}\frac{h_{ab}}{z^2}$,  integration of $h_{ab}$ on $Q$ can also
contribute to the stress tensor on $P$. So $ T_e^{ab}$ and $ T^{ab}$
are different generally. Interestingly, they always yield the same
Weyl anomaly $ T_e{}^{a}_{\ a}=T^a_{\ a}$ due to
$T_{Q}{}^{\alpha}_{\ \alpha}=0$ and the fact that the integration on
$Q$, i.e. $dz z^m$, cannot produce terms of order $O(z^0)$. An
advantage of $T_e^{ab}$ is that it is always finite by definition
$T_e^{ab}=\frac{2}{\sqrt{\sigma_0}}\frac{\delta I_{\rm eff}}{\delta
  \sigma_{0 ab}}$, since 
$I_{\rm eff}$
is finite. The integration of
the other components of  $h_{\alpha\beta}$ on $Q$ give the new
operator $O$ on $P$. It is worth noting that since $h_{\alpha\beta}$
is related to $g_{ij}$ on-shell, the new operator $O$ coming from
$h_{\alpha\beta}$ is also related to geometric quantity derived from
$g^{(0)}_{ij}$. According to \cite{McAvity:1993ue}, such geometric
quantity appears naturally as the new operator on the boundary of
BCFT.

\subsection{4d BCFT}

Now we study the holographic renormalization for 4d BCFT, which is
more subtle. We find that one has to add  squared extrinsic curvature
terms on the corner $P$  in order to make the action finite.\\
We propose the following renormalized action
\begin{eqnarray}\label{regularizedactionAdS5}
I_{\rm ren}&=&\int_N dx^5\sqrt{G} (R-2 \Lambda) + 2\int_Q
dx^4\sqrt{h}( K-T)+ 2\int_M dx^4\sqrt{g}(K-3-\frac{1}{4}R_M
)\nonumber\\
&&+2\int_P dy^3 \sqrt{\sigma}(\theta -\theta_0-\frac{1}{2}K_M+\alpha
R_P+ \beta \text{Tr}\bar{K}^2_Q +\gamma).
\end{eqnarray}
Similar to the case of 3d BCFT, $I_M$ includes the usual counterterms
in holographic renormalization
\cite{Balasubramanian:1999re,deHaro:2000vlm}, $\theta_0=\theta(z=0)$
is a constant \cite{Nozaki:2012qd} and $K_M$ is the
Gibbons-Hawking-York term for $R_M$ on $M$.  It is worth noting that
the induced metric on $Q$ is AdS-like, i.e., it can be rewritten into
the form of eq.(\ref{AdSmetric}) except that now $g_{ij}$ are in
powers of $z$ instead of $z^2$. In spirit of the holographic
renormalization for asymptotically AdS, one can add a constant term
$\gamma$ and an intrinsic curvature term $R_P$ into $I_P$. However,
they are not enough to make the action finite. Instead, we have to add
the extrinsic curvature terms $ \text{Tr}\bar{K}^2_Q$ on $P$. This
maybe due to the presence of the singular corner $P$ and the non-AdS
metric on $Q$. Note that $R_P\sim  \bar{K}^2_Q \sim O(z^2)$ are
designed to delete the $O(\frac{1}{z})$ divergence in the
action \footnote{We have $K_Q{}^a_b \sim O(1)$ and
  $\bar{K}_Q{}^a_b\sim O(z)$. Thus only the combination $\text{Tr}
  \bar{K}^2_Q$ is of order $O(z^2)$.  }. It should be mentioned that
$\bar{K}_{Q\ ab}$  can be regarded as new boundary operator from the
viewpoints of BCFT, since it is defined by the embedding from $P$ to
$Q$ rather than to the spacetime where BCFT lives. On the other hand,
$K_{M\ ab}$ is not an independent operator, since it is defined by the
derivatives of the metric for BCFT. As a result, if we add $K_M^2\sim
O(z^2)$ terms on $P$, we get ill-defined stress tensors with
$\partial_x \delta(x) T_{xa}$,  where $x=0$ denotes the location of
$P$. This means there is energy flowing outside $P$, which is not a
well-defined BCFT. For these reasons, we propose
eq.(\ref{regularizedactionAdS5}) as the renormalized action.

Substituting eqs.(\ref{AdSmetric}-\ref{uniTa}) into the action
(\ref{regularizedactionAdS5}), we can solve $\alpha$, $\beta$ and $\gamma$ that
make a finite action. It is remarkable that $a_3$ and $a_4$ disappear
in the divergent terms of the action (\ref{regularizedactionAdS5})
once we impose the universal relations (\ref{uniTa}).  Thus the
solutions to  $\alpha$, $\beta$ and $\gamma$ are irrelevant to $a_3$ and
$a_4$. After some calculations, we get
\begin{eqnarray}\label{abAdS5}
\alpha=-\frac{1}{4}\sinh \rho_*,\ \  \beta=\frac{1}{4} \cosh\rho_*
\coth\rho_*, 
\ \ \gamma=0.
\end{eqnarray}
A quick way to derive eq.(\ref{abAdS5}) is to consider $AdS_5$ in the
bulk and choose spherical coordinates and cylindrical coordinates on
$M$ for $\alpha$ and $\beta$, respectively. Note that the new
counterterms $ \alpha R_p\sim \beta \text{Tr}\bar{K}^2_Q \sim
O(\rho_*)$ vanish for the trivial boundary condition $\rho_*=0$.

Now we are ready to calculate the boundary contributions to Weyl anomaly.
Similar to the 3d case,  from the constant Weyl transformations
$\sigma_{ab}\to e^{2\epsilon}\sigma_{ab}$, $\sqrt{\sigma}\to
e^{3\epsilon} \sqrt{\sigma}$, $\theta \to \theta$, $K_M\to
e^{-\epsilon} K_M$, $R_P\to e^{-2\epsilon} R_P$ and
$\text{Tr}\bar{K}^2_Q\to e^{-2\epsilon} \text{Tr}\bar{K}^2_Q$, we can
read off the  boundary Weyl anomaly as
\begin{eqnarray}\label{HBWAstress2AdS5}
\int_P dy^3\sqrt{\sigma_0}\left<T^a_a\right>_P=2\int_P
dy^3\sqrt{\sigma} \left( (d-1)(\theta-\theta_0)-K_M +(d-3)(\alpha
R_P+\beta \text{Tr}\bar{K}^2_Q)\right),
\end{eqnarray}
To make eq.(\ref{HBWAstress2AdS5}) finite, we solve
\begin{eqnarray}\label{a3AdS5}
a_3=-\frac{1}{72} \sinh\rho_* \left(\cosh (2 \rho_* )
\left(\mathcal{R}+4 q-k^2-9 \text{Tr}k^2\right)+2 \left(\mathcal{R}+4
q-k^2-6 \text{Tr}k^2\right)\right),
\end{eqnarray}
which is exactly the solution to our proposed boundary condition
(\ref{mixedBCN2}).
Substituting the above $a_3$ into eq.(\ref{HBWAstress2AdS5}), we get
\begin{eqnarray}\label{BWAAdS5}
\left<T^a_a\right>_P&=&\frac{1}{54} \left[-27 \text{sech}^2(\rho_* )
  (48 a_4+qk-6 l-2 k\text{Tr}k^2-6
  \text{Tr}k^3+\text{Tr}(k\mathcal{R})+7 \text{Tr}(kq)) \right. \nonumber\\
&& -3 \cosh (2 \rho_* ) \left(k \left(-3 \mathcal{R}-12 q+k^2+27
  \text{Tr}k^2\right)+27 l+90 \text{Tr}k^3-9
  \text{Tr}(k\mathcal{R})-63 \text{Tr}(kq)  \right)\nonumber\\
&&\left. q +9 k \mathcal{R}+9 kq-81 l-13 k^3+45 k \text{Tr}k^2-54
  \text{Tr}(k\mathcal{R})+54 \text{Tr}(kq)\ \right]
\end{eqnarray}
 Comparing eq.(\ref{a4AdS5}) with eq.(\ref{4dBWA1}), 
we find that they match if and only if
\begin{eqnarray}\label{a4AdS5}
a_4&=&\frac{1}{1728} \big{[} 24  k\mathcal{R}-14 k^3-21 k q+90 k 
\text{Tr}k^2+135 l+108 \text{Tr}k^3-90 \text{Tr}(k\mathcal{R})-144 
\text{Tr}(kq)\nonumber\\
&&+4 \cosh (2 \rho_* ) \left(6 k\mathcal{R}-4 k^3+6 k q-9 
\left(3 l+6 \text{Tr}k^3+\text{Tr}(k\mathcal{R})-5
\text{Tr}(kq)\right)   
\ \right)\nonumber\\
&&+\cosh (4 \rho_* )\left(-2 k^3+9 k q-18 k \text{Tr}k^2-9 
\left(3 l+12 \text{Tr}k^3-2 \text{Tr}(k\mathcal{R})-8 \text{TR}(kq) 
\right )\ \right) \big{]},
\end{eqnarray}
which is again the solution to the boundary condition
(\ref{mixedBCN2}) we proposed. The other choices of boundary
conditions give different $a_3$ and $a_4$  and thus can be
excluded. In the above calculations, we have used the following
formulas
\begin{eqnarray}\label{foumulasbird1}
&&E_4^{\rm bdy}=4 \left(2 \text{Tr}(k\mathcal{R})-k\mathcal{R}
+\frac{1}{3}k^3-k\text{Tr}k^2+\frac{2}{3} \text{Tr}k^3\right),\\ \label{foumulasbird2}
&&\text{Tr}\bar{k}^3=\frac{2}{9} k^3-k\text{Tr}k^2+\text{Tr}k^3,\\ \label{foumulasbird3}
&& C^{ac}_{\ \ \ b c}
\bar{k}_{\ a}^b=-\frac{1}{6}k\mathcal{R}-\frac{1}{6}kq
+\frac{1}{6}k^3-\frac{1}{2}k\text{Tr}k^2+\frac{1}{2}\text{Tr}(k\mathcal{R})
+\frac{1}{2}\text{Tr}(kq)
\end{eqnarray}
in Gauss normal coordinates (\ref{BCFTmetric}). Since the calculations
are quite complicated, the non-patient readers can study some simple
examples instead. For example, $AdS$ in spherical coordinates and
cylindrical coordinates are good enough to reproduce most of the
results above.

To sum up, we have developed a scheme of holographic renormalization
for BCFT. We find that it reproduces the correct boundary Weyl anomaly
eqs.(\ref{3dBWA1},\ref{4dBWA1}) only when $Q$ is determined by
eq. (\ref{mixedBCN2}). This is a non-trivial check of  our proposal
for holographic BCFT.

\section{General Boundary Condition}

In this section, we consider more general boundary conditions for
BCFT. As we have mentioned before, the constant $T$ in the
gravitational action eq.(\ref{action1}) can be regarded as the
holographic dual of boundary conditions for BCFT, since it is closely
related to boundary central charges. Naturally, we propose to add
intrinsic curvature terms on $Q$ to mimic general boundary
conditions. For simplicity, we focus on the case of Ricci scalar. Now
the gravitational action for holographic BCFT becomes
\begin{eqnarray}\label{generalaction}
I=\int_N \sqrt{G} (R-2 \Lambda) + 2\int_Q \sqrt{h} (K-T-\lambda R_Q)
+ 2\int_M \sqrt{g}\ K+2\int_P \sqrt{\sigma}\ \theta,
\end{eqnarray}
where $\lambda$ is a constant. Similarly, we suggest to impose the
mixed boundary conditions on $Q$ with the non-trivial one given by
\begin{eqnarray}\label{generalmixedBCN2}
T_{BY}{}^{\alpha}_{\  \alpha}=2(1-d)K+2dT+2\lambda (d-2)R_Q=0.
\end{eqnarray}
Below we will apply the methods of sect. 3 and sect. 4 to investigate
the boundary contributions to Weyl anomaly. As it is expected, we find
the boundary central charges depend on the new parameter
$\lambda$. And again, these two methods give the same results only if
the bulk boundary $Q$ is determined by the traceless-stress-tensor
condition eq.(\ref{generalmixedBCN2}).

\subsection{General boundary Weyl anomaly I}

Now let us use the method of sect.2 to derive the boundary Weyl
anomaly. For simplicity, we focus on $AdS_4$ with spherical
coordinates and cylindrical coordinates below. 
The generalization to higher dimensions and other metrics is straightforward.
\begin{eqnarray}\label{pureAdSmetric}
ds^2&=&\frac{dz^2+ dr^2+r^2 d\theta^2+r^2\sin^2\theta d\phi^2}{z^2}, 
\ \ \text{spherical coordinates}\\
ds^2&=&\frac{dz^2+ dr^2+r^2 d\theta^2+dy^2}{z^2},\ \ \ \  
\ \ \ \  \ \ \ \ \text{cylindrical coordinates}.\label{pureAdS4metric}
\end{eqnarray}
$P$ is at $r=r_0$ and $Q$ is given by $r=r(z)$ with
\begin{eqnarray}\label{pureAdSQ}
r=r_0+\sinh\rho_* z-\frac{ k}{4}\cosh^2\rho_* z^2+a_3 z^3+\cdots
\end{eqnarray}
where $k$ is $\frac{2}{r_0}$ for sphere and  $\frac{1}{r_0}$ for
cylinder. From the leading term of eq.(\ref{generalmixedBCN2}), we can
re-express $T$ in terms of $\rho_*$ and $\lambda$. In general, we get
\begin{eqnarray}\label{pureAdST}
T=(d-1) \tanh\rho_*+\lambda (d-1) (d-2)\text{sech}^2\rho_*.
\end{eqnarray}
Substituting eqs.(\ref{pureAdSmetric}-\ref{pureAdST}) into the action 
(\ref{generalaction}) and selecting the logarithmic divergent term
after the integral along r and z, we can obtain the boundary Weyl
anomaly. Similarly, one can check that $I_M$ and $I_P$ in the action
(\ref{generalaction}) and $a_3, a_4$ in the embedding function
(\ref{pureAdSQ}) are irrelevant in the above derivations. Rewriting
the final results into covariant form, we obtain
\begin{eqnarray}\label{3dBWA1gen}
\left<T^a_a\right>_P= \sinh \rho_*(1-2\lambda \coth\rho_*)\  
\mathcal{R}-\sinh \rho_*(1-2\lambda \tanh\rho_*)\  \text{Tr} \bar{k}^2.
\end{eqnarray}
Interestingly, now the central charges with respect to $ \mathcal{R}$
and $ \text{Tr} \bar{k}^2$ become independent, which implies that
there are two independent boundary central charges for 3d BCFT
generally. This is the expected result, since every independent Weyl
invariant should correspond to an independent central charge.
The above discussions can be easily generalized to higher dimensions
and general metrics. 
For 4d BCFT, we obtain
\begin{eqnarray}\label{4dBWA1gen}
\left<T^a_a\right>_P= \frac{1}{8} E_4^{\rm bdy}+\left
(\cosh(2\rho_*)(1-4 \lambda  \tanh\rho_*)-\frac{1}{3}\right) \text{Tr}
\bar{k}^3 -\cosh(2\rho_*)(1-4 \lambda  \tanh\rho_*) C^{ac}_{\ \ \ b c}
\bar{k}_{\ a}^b\nonumber\\
\end{eqnarray}
Now the central charges related to $ \text{Tr} \bar{k}^3$ and
$C^{ac}_{\ \ \ b c} \bar{k}_{\ a}^b$ are still not independent. One
can check that, by adding more general curvatures in $I_Q$, the
boundary central charges can indeed become independent. 
For example, let us consider the action
\begin{eqnarray}\label{generalactionRRR}
I=\int_N \sqrt{G} (R-2 \Lambda) + 2\int_Q \sqrt{h} (K-T-\lambda R_Q-\lambda_2 \bar{R}_{Q}{}^{\alpha}_{\beta} \bar{R}_{Q}{}^{\beta}_{\gamma} \bar{R}_{Q}{}^{\gamma}_{\alpha} )
+ 2\int_M \sqrt{g}\ K+2\int_P \sqrt{\sigma}\ \theta,
\end{eqnarray}
where $\bar{R}_{Q}{}^{\alpha}_{\beta}=R_{Q}{}^{\alpha}_{\beta}+\frac{(d-1)}{\cosh^2\rho_*}\delta^{\alpha}_{\beta}$. Following the above approach, we derive
\begin{eqnarray}\label{4dBWA1genRRR}
\left<T^a_a\right>_P&=& \frac{1}{8} E_4^{\rm bdy}+\left
(\cosh(2\rho_*)(1-4 \lambda  \tanh\rho_*)-\frac{1}{3}-16\lambda_2  \tanh ^3\rho_* \text{sech}^2\rho_*\right) \text{Tr}
\bar{k}^3\nonumber\\
&& -\cosh(2\rho_*)(1-4 \lambda  \tanh\rho_*) C^{ac}_{\ \ \ b c}
\bar{k}_{\ a}^b
\end{eqnarray}
We remark that in obtaining the results \eq{4dBWA1gen} and \eq{4dBWA1genRRR}, it is necessary to
consider non-AdS solutions in order
to derive the central charge related to $C^{ac}_{\ \ \ b c}
\bar{k}_{\ a}^b$ since  $C_{abcd}=0$ for $AdS$.

\subsection{General boundary Weyl anomaly II}

In this section, we take the method of sect.3 to study the boundary 
Weyl anomaly for general boundary conditions. Due to the Ricci scalar
in $I_Q$ (\ref{generalaction}), we should add new a
Gibbons-Hawking-York term $K_Q$ in $I_P$. Recall that the induced
metric on $Q$ is AdS-like, i.e., it can be rewritten into the form of
eq.(\ref{AdSmetric}) except that now $g_{ij}$ are in powers of $z$
instead of $z^2$. In spirit of the holographic renormalization for
asymptotically AdS, one can add a constant term and intrinsic
curvature terms on $P$. Besides, from the experience of sect.3, one
has to add extrinsic curvature terms in order to make the action
finite generally. This is may because of the presence of the corner
$P$ and the non-AdS metric on $Q$.  Based on the above discussions, we
propose the following renormalized action for 3d and 4d BCFT
\begin{eqnarray}\label{regularizedactiongen}
I_{re}&=&\int_N dx^{d+1}\sqrt{G} (R-2 \Lambda) + 2\int_Q dx^d\sqrt{h}(
K-T-\lambda R_Q)+ 2\int_M dx^d\sqrt{g}(K-(d-1)-\frac{1}{2(d-2)}R_M
)\nonumber\\
&&+2\int_P dy^{d-1} \sqrt{\sigma}(\theta
-\theta_0-\frac{1}{d-2}K_M-2\lambda K_Q+\alpha R_P+ \beta
\text{Tr}\bar{K}^2_Q+\gamma ).
\end{eqnarray}
where $\alpha,\beta,\gamma$ are parameters and will 
be determined below. For 3d BCFT, we have $\gamma=2\lambda
\text{sech}\rho_*$, and $\alpha,\gamma$ are free parameters since they
are related to finite counterterms. Below, we focus on 4d BCFT.

For simplicity, we focus on $AdS$ with spherical coordinates and
cylindrical coordinates.
\begin{eqnarray}\label{pureAdS5metric1}
ds^2&=&\frac{dz^2+ dr^2+r^2
  d\Omega^2}{z^2},\ \  \ \ \ \ \ \ \ \ \ \ \  \ \ \ \  \ \ \ \ \ \ \text{spherical
  coordinates},\\
ds^2&=&\frac{dz^2+ dr^2+r^2 d\theta^2+\sin^2\theta
  d\phi^2+dy_2^2}{z^2},\ \text{cylindrical coordinates
  I}\label{pureAdS5metric2},\\
ds^2&=&\frac{dz^2+ dr^2+r^2
  d\theta^2+dy_1^2+dy_2^2}{z^2},\ \ \ \ \ \ \ \ \text{cylindrical
  coordinates II},\label{pureAdS5metric3}
\end{eqnarray}
Again, we put $P$ at $r=r_0$ and label $Q$ by $r=r(z)$ with
\begin{eqnarray}\label{pureAdS5Q}
r=r_0+\sinh\rho_* z-\frac{k}{6}\cosh^2\rho_* z^2+a_3 z^3+a_4 z^4+\cdots
\end{eqnarray}
where $k$ take values $(\frac{3}{r_0}, \frac{2}{r_0}, \frac{1}{r_0})$ 
for the metrics
(\ref{pureAdS5metric1},\ref{pureAdS5metric2},\ref{pureAdS5metric1}),
respectively. From the traceless-stress-tensor condition
eq.(\ref{generalmixedBCN2}), we can solve the above embedding
function. For the spherical metric (\ref{pureAdS5metric1}), we can get
exact solution
\begin{eqnarray}\label{sphereADSr}
r=\sqrt{r_0^2\cosh^2\rho_*-(z-r_0 \sinh\rho_*)^2}.
\end{eqnarray}
For the first kind of cylindrical metric eq.(\ref{pureAdS5metric2}), we obtain
\begin{eqnarray}\label{sphereADSIa3}
a_3&=&\frac{\cosh\rho_*  (9 \sinh (2 \rho_* )-4 \lambda  (9 \cosh (2
  \rho_* )-8))}{54 r_0^2 (1-4 \lambda  \tanh\rho_*)},\\
a_4&=&\frac{\cosh\rho_* (-84 \lambda  \sinh\rho_*+44 \lambda  \sinh (3
  \rho_* )+\cosh\rho_*-11 \cosh (3 \rho_* ))}{108 r_0^3 (1-4 \lambda
  \tanh\rho_* )}\label{sphereADSIa4}
\end{eqnarray}
As for the second second of cylindrical metric
eq.(\ref{pureAdS5metric3}), we have
\begin{eqnarray}\label{sphereADSIIa3}
a_3&=&\frac{\cosh\rho_*(9 \sinh (2 \rho_* )-36 \lambda  \cosh (2
  \rho_* ) +28 \lambda )}{108 r_0^2 (1-4 \lambda  \tanh\rho_*)},\\
a_4&=&\frac{\cosh\rho_* (4 \lambda  (47 \sinh (3 \rho_*)-81 \sinh
  (\rho_* ))+19 \cosh (\rho_* )-47 \cosh (3 \rho_* ))}{864 r_0^3 (1-4
  \lambda  \tanh\rho_*)}\label{sphereADSIIa4}
\end{eqnarray}

Substituting eqs.(\ref{pureAdS5metric1}-\ref{pureAdS5Q}) into the
action (\ref{regularizedactiongen}) and requiring the action finite,
we derive
\begin{eqnarray}\label{ablsolution}
\alpha=\lambda \cosh\rho_*-\frac{\sinh\rho_*}{4}
,\ \ \beta=\frac{1}{4} \cosh\rho_* \coth\rho_*-\lambda
\cosh\rho_*,\ \ \gamma=4 \lambda\  \text{sech}\rho_*.
\end{eqnarray}
Again, $a_3$ and $a_4$ do not appear in the divergent terms of the
action (\ref{regularizedactiongen}). Actually, we can use only two of
the three examples in
eqs.(\ref{pureAdS5metric1},\ref{pureAdS5metric2},\ref{pureAdS5metric3})
to derive eq.(\ref{ablsolution}). The third one provides a double
check of our calculations.

Now we are ready to calculate the boundary contributions to Weyl
anomaly. Similar to the cases of sect. 3,  with the help of constant
Weyl transformations, we can read off the  boundary Weyl anomaly as
\begin{eqnarray}\label{genBWAAdS5}
&&\int_P dy^3\sqrt{\sigma_0}\left<T^a_a\right>_P\nonumber\\
&=&2\int_P dy^3\sqrt{\sigma} \left( d\,
  \lambda+(d-1)(\theta-\theta_0)-K_M-2\lambda(d-2)K_Q +(d-3)(\alpha
  R_P+\beta \text{Tr}\bar{K}^2_Q)\right),
\end{eqnarray}
Substituting eqs.(\ref{pureAdS5metric1}-\ref{ablsolution}) into the
above formula, we can derive the boundary Weyl anomaly for the three
examples in
eqs.(\ref{pureAdS5metric1},\ref{pureAdS5metric2},\ref{pureAdS5metric3}),
which exactly agrees with the result eq.(\ref{4dBWA1gen}) of last
subsection. This is a strong support to our proposal of holographic
BCFT with zero trace of the stress tensors on $Q$, i.e.,
$T_{BY}{}^{\alpha}_{\ \alpha}|_Q=0$.

\section{Holographic Entanglement Entropy}

\subsection{General formula}

Let us go on to discuss the holographic entanglement entropy.
Following \cite{Ryu:2006bv,Lewkowycz:2013nqa}, it is not difficult to
derive the holographic entanglement entropy
for a $d$-dimensional
BCFT, which is also
given by the area of  minimal surface
 \begin{eqnarray}\label{HEE}
S_A=\frac{\text{Area}(\gamma_A)}{4 G_N},
\end{eqnarray}
where $A$ is a $(d-1)$-dimensional
subsystem on $M$, and $\gamma_A$ denotes the minimal
surface which ends on $\partial A$.   What is new for BCFT is that the
minimal surface could also end on the bulk boundary $Q$, when the
subsystem $A$  is close to the boundary $P$. See Fig.1 for example.

We could keep the endpoints of extreme
surfaces $\gamma_A'$ freely on $Q$, and select the one with minimal
 area as $\gamma_A$. It follows that
$\gamma_A$ is orthogonal to the boundary $Q$ when they
intersect
 \begin{eqnarray}\label{normalAQ}
n^a_{\gamma_A} \cdot n_Q|_{\gamma_A\cap Q}=0.
\end{eqnarray}
Here $n_Q$ is the normal vector of $Q$ and
$n^a_{\gamma_A}$ are the two independent normal vectors of
$\gamma_A$.
It is easy to see that if $\gamma'_A$ is not normal to $Q$, one can 
always deform $\gamma'_A$ to decrease the area until it is normal to
$Q$ \footnote{We thank Dong for 
emphasizing this point to us.}. Let
us take an example in  Fig.1 to illustrate this. For simplicity, we
focus on static spacetime and constant time slice. Then the normal
vector of
$\gamma_A$ alone time is orthogonal to $n_Q$ trivially. It is worth
keeping in mind that the induced metric on constant time slice is
Euclidean and positive definite. Below we focus on the case $d=2$. It
is straightforward to generalize our discussions to higher dimensions.
Consider an extreme surfaces $OA$ in Fig.1, where $O$ is a fixed point
in the bulk, and $OA$ is not normal to the boundary $Q$. Then select
an arbitrary point $B$ alone $OA$ as long as it is near enough to the
boundary. Starting from $B$, we can construct a minimal surface $BA'$
that is normal to the boundary and ending on the boundary at $A'$
. Since the metric is positive definite and $B$ is near enough to the
boundary, we have $BA > BA'$ and thus $OBA > OBA'$. Next we construct
a minimal surface $OA'$ linking $A'$ and $O$. By definition, it is
smaller than $OBA'$. As a result, we have  $OBA > OBA'>OA'$. If $OA'$
is not orthogonal to $Q$ either, we can repeat the above approach
again and again until the extreme surface is normal to $Q$. Now it is
clear that the minimal area condition leads to the orthogonal
condition (\ref{normalAQ}).

\begin{figure}
\centering
\includegraphics[width=8cm]{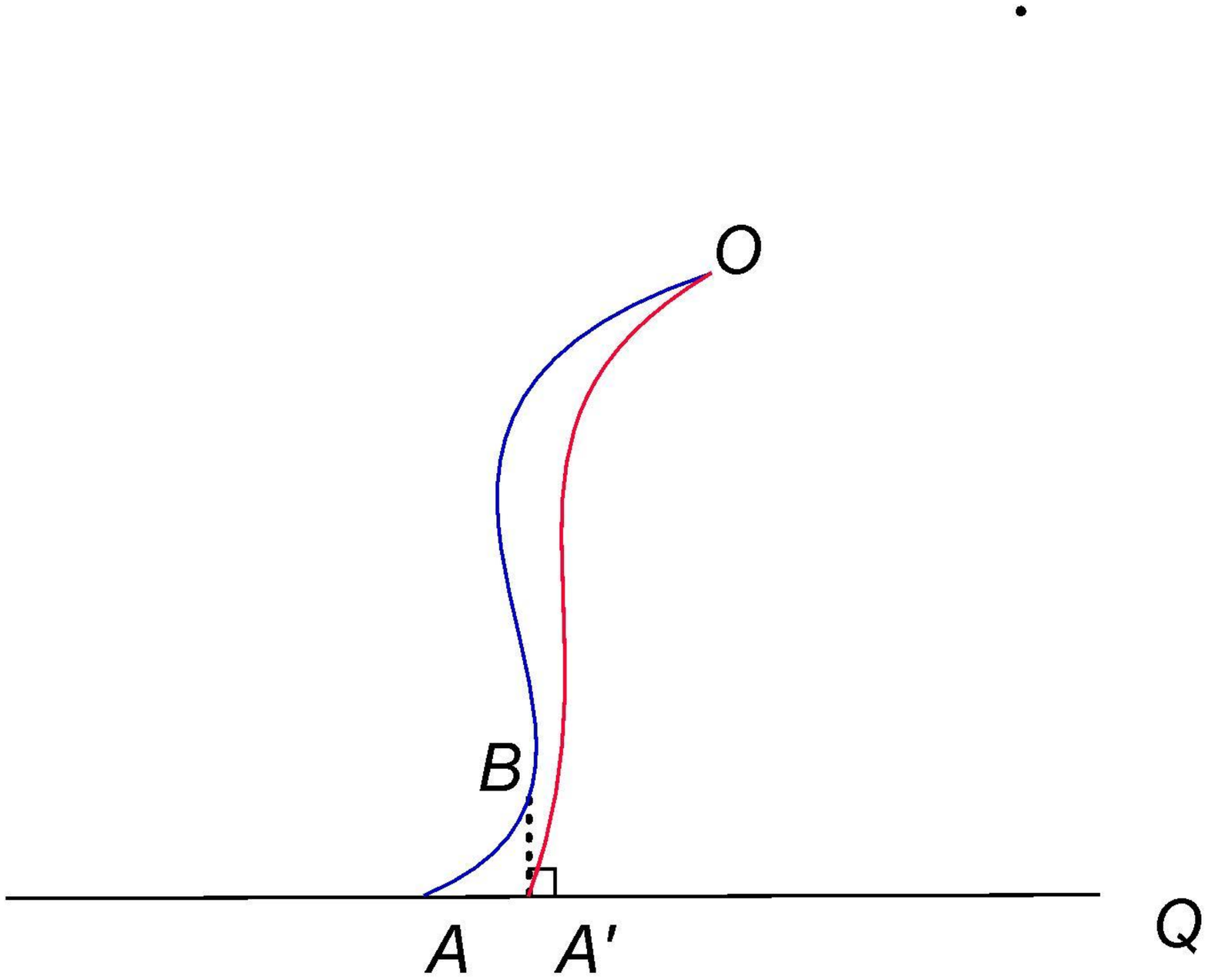}
\caption{$OBA > OBA' >OA'$}
\end{figure}

Another way to obtain the orthogonal condition is that,
otherwise there will arise problems in the holographic
 derivations of entanglement entropy by using the replica trick. In
 the replica method, one considers the $n$-fold cover $M_n$ of $M$ and
 then extends it to the bulk as $N_n$. It is important that $N_n$ is a
 smooth bulk solution. As a result, Einstein equation should be
 smooth on
surface $\gamma_A$.
Now the metric near  $\gamma_A$
is given by \cite{Lewkowycz:2013nqa}
 \begin{eqnarray}\label{minimalsurface}
ds^2=\frac{1}{r^{2\varepsilon}}(dr^2+r^2 d\tau^2)+
\left(g_{ij}+2\mathcal{K}_{aij}x^a+ O(r^2) \right)dy^i
dy^j, \nonumber
\end{eqnarray}
where
$\varepsilon \equiv 1-\frac{1}{n}$, $r$ is coordinate normal to the
surface, $\tau\sim \tau+2\pi n$ is the Euclidean time, $y^i$ are 
coordinates along the surface,
$x^a =(r \cos \tau, r \sin \tau)$ and
$\mathcal{K}_{aij}$ are the two extrinsic
curvature tensors. Going to complex
coordinates $z=r e^{i\tau}$, the $zz$ component of Einstein equations
 \begin{eqnarray}\label{Eineq}
R_{zz}=-\mathcal{K}_z \frac{\varepsilon}{z}+ \cdots
\end{eqnarray}
is divergent unless the trace of extrinsic curvatures vanish
$\mathcal{K}_a=0$. This gives
the condition for a minimal
surface \cite{Lewkowycz:2013nqa}.
Labeling the boundary $Q$ by $f(z,\bar{z},y)=0$, we obtain
the extrinsic curvature of $Q$ as
 \begin{eqnarray}\label{BCregular}
K\sim \varepsilon \ \partial_z f \partial_{\bar{z}} f (\frac{ \partial_z
  f }{\bar{z}}+\frac{ \partial_{\bar{z}} f }{z} )+ \cdots .
\end{eqnarray}
So the boundary condition (\ref{mixedBCN2}) is smooth only if
$\partial_z f|_{\gamma_A\cap Q} = \partial_{\bar{z}} f|_{\gamma_A\cap
  Q}=0 $, which is exactly the
orthogonal condition (\ref{normalAQ}). It should be mentioned that the
smooth requirement of the general boundary conditions
(\ref{mixedBCN1}) may yield more constraints in addition to the
orthogonal condition (\ref{normalAQ}). When $A_{\alpha\beta}$ includes
higher curvature terms, sometimes the smooth requirement even leads to
contradictions. This can also help us to exclude a large class of
$A_{\alpha\beta}$ in the boundary condition (\ref{mixedBCN1}).  Since
our boundary condition (\ref{mixedBCN2}) yields the expected
orthogonal condition (\ref{normalAQ}), this is also a support to our
proposal.

In summary, the holographic entanglement entropy for BCFT is
given by RT
formula (\ref{HEE}) together with the orthogonal condition
(\ref{normalAQ}). As we will show below, there appear many new
interesting properties for entanglement due to the presence of
boundaries.

\subsection{Boundary effects on entanglement}

Let us take an simple example to illustrate the boundary effects on
entanglement entropy. Consider Poincare metric of $AdS_3$
 \begin{eqnarray}\label{AdS3}
ds^2=\frac{dz^2+dx^2-dt^2}{z^2},
\end{eqnarray}
where $P$ is at $x=0$. Solving eq.(\ref{mixedBCN2}) for $Q$, we get
$x= \sinh(\rho)  z$ and $T=\tanh \rho \ge 0$. We choose $A$ as an
interval with two endpoints at $x=d$ and $x=d+2l$. Due to the presence
of boundary, now there are two kinds of minimal surfaces, one ends on
$Q$ and the other one does not. It depends on the distance $d$ that
which one has smaller area.
 From eqs.(\ref{HEE},\ref{normalAQ}), we obtain
\be \label{SA}
S_A= \begin{cases}
\frac{1}{2 G_N}\log (\frac{2l}{\epsilon }), &  
d\ge d_c,\\
\frac{\rho}{2 G_N}+\frac{1}{4 G_N}\log 
\Big( \frac{4d(d+2l)}
{\epsilon ^2}
\Big), 
& d \le
d_c,
 \end{cases}
\ee
where $d_c=l\sqrt{e^{-2 \rho }+1}-l$ is the critical distance. The
parameter $\rho$ can be regarded as the holographic dual of the
boundary condition of BCFT, since it affects the boundary entropy
\cite{Takayanagi:2011zk} and the boundary central charges
(\ref{3dBWA1},\ref{4dBWA1}) as the boundary condition does. It is
remarkable that entanglement entropy (\ref{SA}) depends on the
distance $d$ and boundary condition $\rho$ when it is close enough to
the boundary. This is the expected property from the viewpoint of
BCFT, where the correlation functions depend on the distance to the
boundary \cite{McAvity:1993ue}.

To extract the effects of boundary, let us define a new physical
quantity when $A\cap P=0$
\begin{eqnarray}\label{muinformation}
I_A= S^{CFT}_A-S^{BCFT}_A,
\end{eqnarray}
where $S^{CFT}_A$ is the entanglement entropy when the boundary
disappears or is at infinity. For simplicity, we focus on the case
$\rho_*\ge 0$. In the holographic language, $S^{CFT}_A$ is given by
the area of minimal surface that does not end on $Q$. Thus,
$S^{CFT}_A$ is equal to or bigger than $S^{BCFT}_A$ and $I_A$ is
always non-negative. It is expected that boundary does not affect the
divergent parts of entanglement entropy when $A\cap P=0$, so all the
divergence cancel in eq.(\ref{muinformation}). As a result, $I_A$ is
not only non-negative but also finite. For the example discussed above, we
have
\begin{eqnarray}\label{IA}
I_A=\begin{cases}
0, & d \ge d_c\\
\frac{1}{4 G}\log (\frac{l^2}{d (d+2 l)})-\frac{\rho }{2 G}, & 0<d < d_c,
 \end{cases}
\end{eqnarray}
which is indeed both non-negative and finite. Actually in this simple
example, $I_A$ is just one half of the mutual information between $A$
and its mirror image, so it must be non-negative and finite. See Fig.1. for
example. For this simple case, the metric at the mirror image $O'$ of
a point $O$ is given by the metric at the point $O$. One should keep
in mind that the mirror image is only an auxiliary tool, there is no
real spacetime outside the boundary $Q$.
\begin{figure}
\centering
\includegraphics[width=8cm]{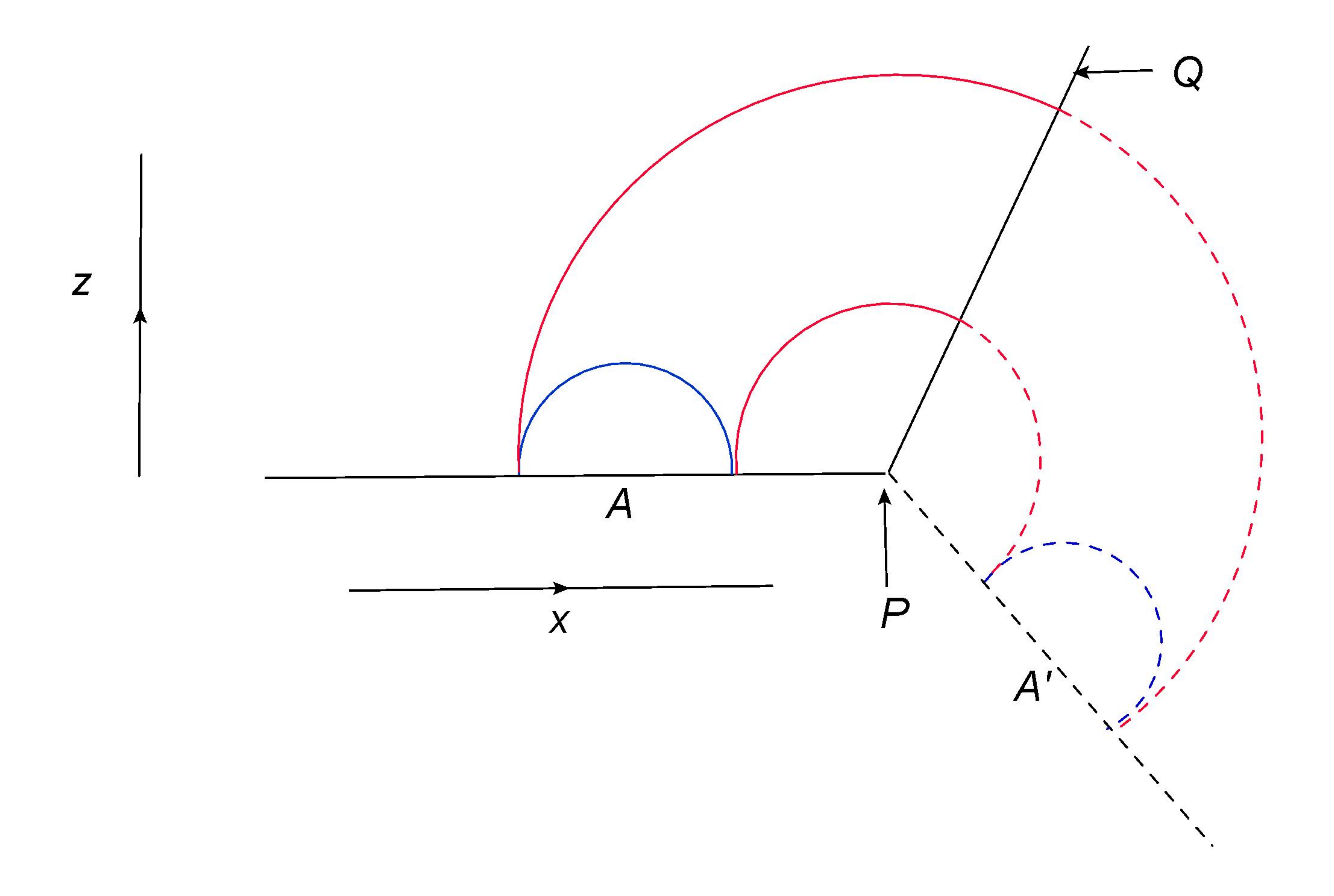}
\caption{Subsystem $A$ and its mirror image $A'$}
\end{figure}

\subsection{Entanglement entropy for stripe}
In this subsection, we study the entanglement entropy of stripe in
general dimensions.
Consider a BCFT defined on $M$ the half space $x\equiv x_1<0$, and
consider a subsystem
$A$ given by the constant time slice $-l<x_1<0$, $-L<x_2, x_3, \cdots,
x_{d-1}<L$. The bulk boundary $Q$ is given by the co-dimension-1 surface $x=z
\tan\theta$.
Here  the parameter
  $\theta=\arctan(\sinh(\rho_*))$, where $\rho_*$ is the parameters
  that we used in previous section. $\theta+\frac{\pi}{2}$ is the
  angle between $Q$ and $M$. 
See Fig. 3.
\begin{figure}\label{stripe}
\centering
\includegraphics[width=10cm]{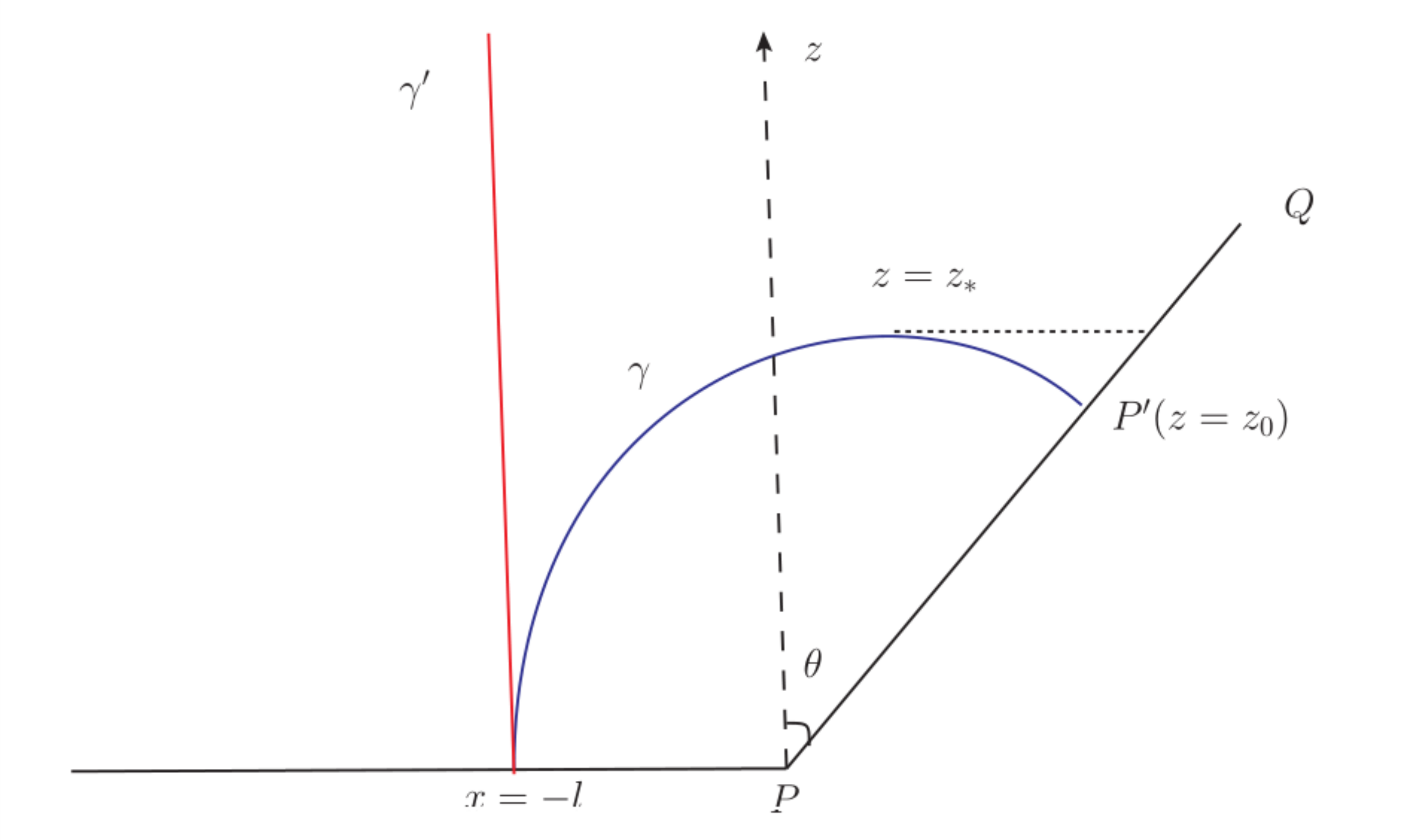}
\caption{The minimal surface of stripe.}
\end{figure}
~\\

Let the minimal surface $\gamma$ be specified by the equation $x=x(z)$ with the
boundary condition 
\be\label{bc0}
 x(0)=l. 
\ee
The induced metric on the minimal surface $\gamma$ is
\begin{eqnarray}
ds^2_\gamma=\frac{(1+x'(z))dz^2+\sum_{i=2}^{d-1} dx_i^2}{z^2}.
\end{eqnarray}
and gives  the equation of motion 
\begin{eqnarray}\label{MinimalEOMstrip}
\frac{x'(z)}{z^{d-1}\sqrt{1+x'(z)^2}}=C, \quad\mbox{$C=$ constant},
\end{eqnarray}
for the minimal surface. There are two kinds  minimal surface. 
If $C=0$, the solution is $x=-l$. The other situation is $C\neq 0$, in
this case, assume when $z=z_*$, $x'(z_*)=\infty$. Let $(x_0,z_0)$ be
the coordinates of  the point $P'$ where
$Q$ and $\gamma$
intersect. It is $x_0\equiv x(z_0)=z_0\tan \theta$.  Denote the
unit normal vectors of $Q$ by $n^Q_\mu$, the unit normal vector of
$\gamma$ by $n^\gamma_\mu$. At point $P'$ we have $n^Q\cdot
n^\gamma=0$. This gives the boundary condition
\begin{eqnarray}\label{condition1}
x'(z_0)=-\cot\theta.
\end{eqnarray} 
Now we solve \eq{MinimalEOMstrip} together with the boundary conditions \eq{bc0},
\eq{condition1}. 
Using the condition (\ref{condition1}) we have
\begin{eqnarray}
z_0^{d-1}=z_*^{d-1} \cos\theta,
\end{eqnarray}
and
\begin{eqnarray}
x'(z)=\frac{(\frac{z}{z_*})^{d-1}}{\sqrt{1-(\frac{z}{z_*})^{2(d-1)}}}.
\end{eqnarray}
We also have the relation
\begin{eqnarray}
x_0+l=\int_0^{z_*}x'(z)dz+\int^{z_*}_{z_0}x'(z) dz.
\end{eqnarray}
This allow us to solve for $z_*$,
\begin{eqnarray}\label{zstar}
z_*=\frac{l}{F(\theta)},
\end{eqnarray}
where
\begin{eqnarray}
F(\theta)&=&\int_0^1\frac{x^{d-1}}{\sqrt{1-x^{2(d-1)}}}dx
+\int^1_{(\cos\theta)^{\frac{1}{d-1}}}\frac{x^{d-1}}{\sqrt{1-x^{2(d-1)}}}dx
-\tan \theta (\cos\theta)^{\frac{1}{d-1}}\nonumber \\
&=&\frac{B\left(\cos^2\theta;\frac{d}{2
      (d-1)},\frac{1}{2}\right)}{2(d-1)}+2\frac{\sqrt{\pi }
  \Gamma\left(\frac{d}{2(d-1)}\right)}{\Gamma \left(\frac{1}{2
      (d-1)} \right)}-\tan \theta (\cos\theta)^{\frac{1}{d-1}},
\end{eqnarray}
where $B(x;a,b)$ is the incomplete beta function. 
When $d\ge 3$, there always exist some critical point $\theta_c$  such that
$F(\theta_c)=0$, as we can see in the Fig.2 for $d=3$ and $d=4$. One
can also show that $\theta_c$ is a monotone decreasing function of
$d$. In particular in the limit $d\to +\infty$, $\theta_c\to 0$.

In the limit $\theta\to \theta_c$, $z_*\to +\infty$, the solution will
tend to the case $C=0$, i.e. the solution $x=-l$. 
For $\theta >\theta_c$ there is
only one solution of minimal surface $x=-l$. For $\theta<\theta_c$ we
have two minimal surface solutions, the desired solution is the one
with a smaller area.  
Consider first the surface  $x=-l$. It is easy
to obtain its area 
\be
A_1=\frac{A_0}{(d-2)\epsilon^{d-2}},
\ee
where $A_0$ is the area of the
entangling surface.
\begin{figure}
\centering
\includegraphics[width=7cm]{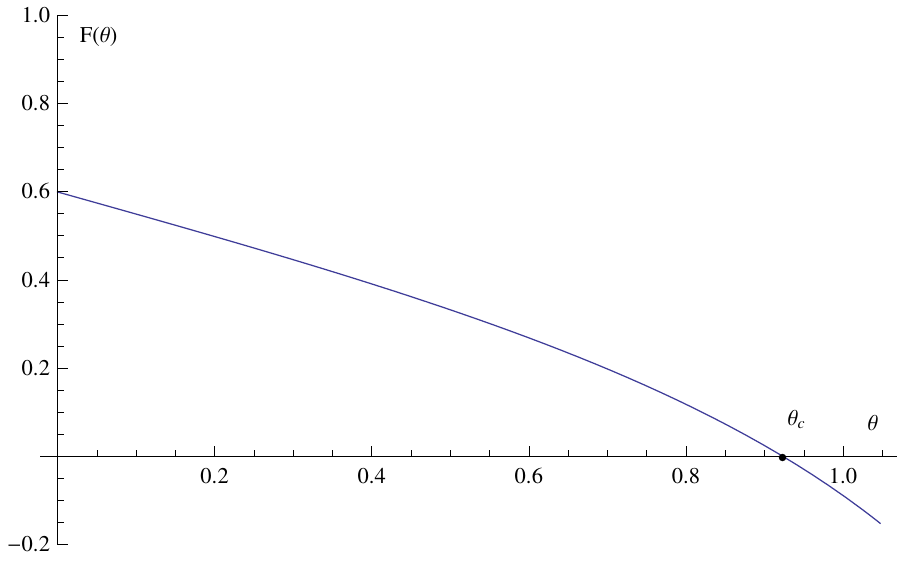}
\includegraphics[width=7cm]{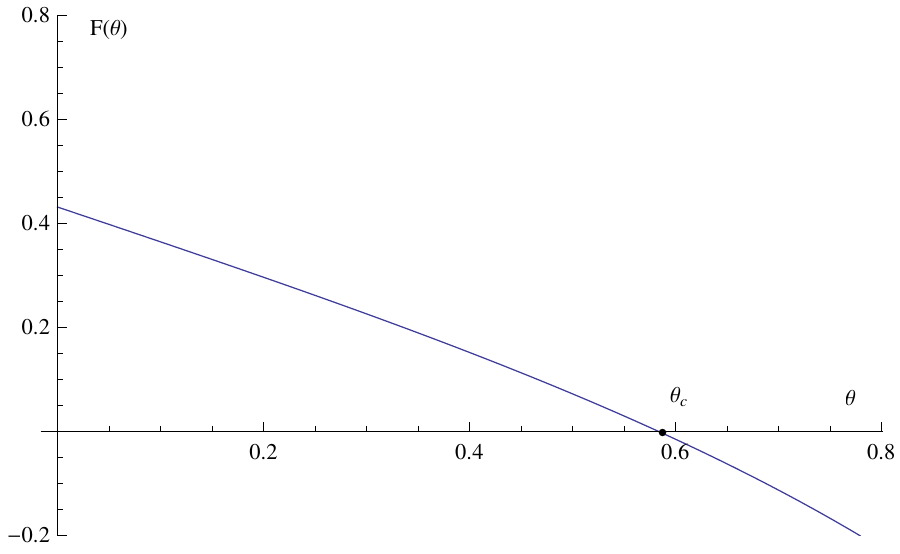}
\caption{$F(\theta)$ for $d=3$(left) and $d=4$(right)}
\end{figure}\label{figurestrip}
The area of the other surface is
\begin{eqnarray}
A_2&=&A_0(\frac{1}{z_*})^{d-2}\Big(\int_{\epsilon/z_*}^1
\frac{dx}{x^{d-1}\sqrt{1-x^{2(d-1)}}}+\int^1_{(\cos\theta)^{\frac{1}{d-1}}}
\frac{dx}{x^{d-1}\sqrt{1-x^{2(d-1)}}}\Big)
\nonumber \\
&=&\frac{A_0}{(d-2)\epsilon ^{d-2}}+
\frac{A_0}{z_*{}^{d-2}}\left(\frac{\sqrt{\pi } \Gamma\big(\frac{-(d-2)}{2
      (d-1)} \big)}{(d-1) \Gamma\big(\frac{1}{2
      (d-1)}\big)}+
\frac{(\cos\theta)^{-\frac{d-2}{d-1}}}{d-2}
  ~_2F_1 \Big(\frac{1}{2},\frac{-(d-2)}{2 (d-1)};\frac{d}{2
      (d-1)};\cos^2\theta\Big)\right)+\cdots\nonumber. \\
~
\end{eqnarray}
Here $\epsilon$ is the cutoff and $_2F_1(a,b;c;z)$ is the hypergeometric
functions. 
In the limit $\theta\to \theta_c$, $z_*\to +\infty$, as a result,
$A_2\to A_1$. One could also show $A_2$ is a monotone decreasing
function of $\theta$ when $\theta<\theta_c$.  Therefore
in the region $\theta<\theta_c$, $A_2<A_1$, the entanglement entropy
is given by $\frac{A_2}{4G}$.  

We remark that our holographic calculation suggests that there is a 
phase transition
at the critical value $\theta=\theta_c$. In our example we see
that $\theta_c$ is independent of the size of the stripe $l$. But it
is probably related to the shape of the entangling surface
in general.  As the
parameter $\theta$ is expected to be dual to the boundary condition of
BCFT, it is interesting to explore what is the nature of the boundary
condition in the field theory that would lead to 
this phase transition in the BCFT.

\section{Entanglement Wedge}

According to \cite{Jafferis:2015del,Dong:2016eik}, a sub-region $A$ on
the AdS boundary is dual to 
an entanglement wedge $\mathcal{E}_A$ in the bulk where all the bulk
operators within $\mathcal{E}_A$ can be reconstructed by using only
the operators of $A$. The entanglement wedge is defined as the bulk
domain of dependence of any achronal bulk surface between the minimal
surface $\gamma_A$ and the  subsystem $A$. Apparently, it seems to
conflict with the holographic proposal of BCFT by
\cite{Takayanagi:2011zk} and us, where the holographic dual of $A$ is
given by $N$, which is larger than $\mathcal{E}_A$ generally. Of
course, there is no contradiction. That is because CFT and BCFT are
completely different theories. For CFT, although we do not know the
information outside, there still exists spacetime outside $A$. As for
BCFT, there is no spacetime outside $A$ at all. Besides, we should
impose suitable boundary conditions for BCFT, while there is no need
to set boundary condition on the entangling surface for CFT.

It is interesting to study the entanglement wedge in the framework of
AdS/BCFT. For simplicity, we focus on the static spacetime and
constant time slice. Recall that the entanglement wedge is given by
the region between the minimal surface $\gamma_A$ and the subsystem
$A$ on $M$. A key observation is that entanglement wedge behaves a
phase transition and becomes much larger than that within AdS/CFT,
when $A$ is increasing and approaching to the boundary.  See Fig.3 for
example. This  phase transition is important for the self-consistency
of holographic BCFT. If there is no phase transition, then the
entanglement wedge is always given by the first kind (left hand side
of Fig.3). When $A$ fills with the whole boundary $M$ and $P$, there
are still large space left outside the entanglement wedge, which means
there are operators in the bulk cannot be reconstructed by all the
operators on the boundary. Thanks to the phase transition, the
entanglement wedge for large A is given by the second kind ( right
hand side of Fig.3). As a result all the bulk operators can be
reconstructed by using the operators on the boundary.
\begin{figure}\label{Wedge}
\centering
\includegraphics[width=8cm]{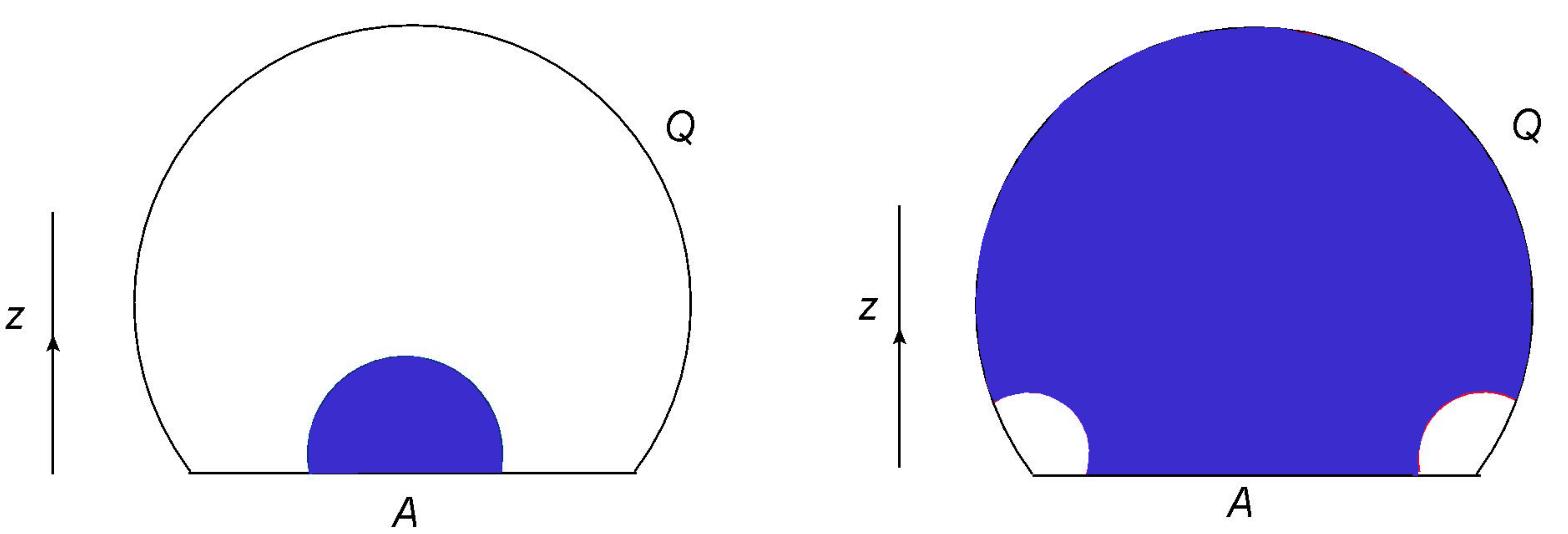}
\caption{Entanglement wedge for small $A$ and large $A$}
\end{figure}
\newpage

\section{Conclusions and Discussions}

In this letter, we
have proposed a new holographic dual of BCFT, which
can accommodate all possible shapes of the boundary $P$
with a unified prescription.  The key idea is
to impose the mixed boundary condition \eq{mixedBCN2} so that there is only one
constraint for the co-dimension one boundary $Q$.
In general there could be
more than one self-consistent boundary
conditions for a theory \cite{Song:2016pwx}, so the
proposals of \cite{Takayanagi:2011zk} and ours have no
contradiction in principle. However, the proposal of
\cite{Takayanagi:2011zk} is too
restrictive to include the general BCFT.
The main advantage of our proposal is that we can deal
with all shapes of the boundary
$P$ easily
and that it can accommodate nontrivial boundary Weyl anomaly as is needed
in a general BCFT.
It is appealing that the
bulk boundary $Q$ is given by a constant mean curvature surface, which
is a natural generalization of the minimal surface.

Applying the new AdS/BCFT, we obtain the expected boundary Weyl
anomaly for 3d and 4d BCFT and
the obtained boundary central charges satisfy naturally a c-like
theorem holographically.
As a by-product, we give a holographic disproof of the 
proposal
\cite{Solodukhin:2015eca} and clarify that the validity of the $S_{RE}=S_{EE}$
conjecture \cite{Herzog:2016kno}
which is based on \cite{Solodukhin:2015eca}
and is sensitively dependent
on the choices of boundary conditions of non-free BCFT.
Besides, we find the holographic entanglement entropy is given by
the RT formula
together with the condition
that the minimal surface must be orthogonal to $Q$ if they intersect.
The presence of boundaries lead to many interesting effects,
e.g. phase transition of the entanglement wedge.
Of course, many things are
left to be explored, for instance, the holographic R\'enyi entropy \cite{
Hung:2011nu,Dong:2016fnf}, the edge modes
\cite{Donnelly:2014fua,Huang:2014pfa}, the shape dependence of
entanglement \cite{Bueno:2015rda,Mezei:2014zla}, the
applications to condensed matter and
the relation between BCFT and quantum information
\cite{Numasawa:2016emc}.
Finally, it is straightforward to
generalize our work to Lovelock gravity, higher dimensions and general
boundary conditions.

\section*{Acknowledgements}
We would like to thank 
X. Dong, L.Y. Hung, F.L. Lin
for useful discussions and comments.
 This work is supported in part by the National
Center of Theoretical Science (NCTS) and the grant MOST
105-2811-M-007-021 of the Ministry of
Science and Technology of Taiwan.

\appendix

\section{Another Derivation of (\ref{finhat2}) }

In sect. 2.1, we have obtained the key result \eq{finhat2} from the PBH
transformation together with the explicit requirement  of covariance under the
residual diffeomorphism of the  gauge fixing condition
(\ref{STgauge}). 
In this appendix, we derive eq.(\ref{finhat2}) directly from the covariant
equation (\ref{mixedBCN2}) and the  gauge fixing (\ref{STgauge}). 
The analysis is manifestly covariant with respect to 
\eq{STgauge} and provides an independent
derivation of the \eq{finhat2}.

To compute $K$, we note that the extrinsic curvature $K_{\alpha \beta}$ on $Q$ is
\begin{eqnarray}
K_{\alpha \beta}=n^Q_\mu K^{\mu}_{\alpha \beta},
\end{eqnarray}
where
\begin{eqnarray}
K^{\mu}_{\alpha \beta}=\partial_\alpha \partial_\beta
X^{\mu}-\gamma_{\alpha\beta}^{\delta}\partial_\delta
X^{\mu}+\Gamma_{\nu\lambda}^{\mu}\partial_\alpha
X^{\lambda}\partial_\beta X^{\nu},
\end{eqnarray}
$\gamma_{\alpha\beta}^{\delta}$ is the Christoffel symbol for the
induced metric $h_{\a\b}$ 
and
$n^Q_{\mu}$ is the unit normal vector on $Q$.
The components of $K^{\mu}_{\alpha \beta}$ can be worked out easily.
Expanded in powers of $\sqrt{\tau}$ for small $\tau$, we have
\begin{eqnarray}
&&K^\rho_{\tau\tau}=\frac{(\xfu)^2}{2\tau}-\frac{1}{\sqrt{\tau}}\frac{\xfu
    \xsd +\frac{1}{2}\gaterm}{1+(\xfu)^2}+\frac{2}{\sqrt{\tau}}\xfu
  \xsd+\frac{1}{\sqrt{\tau}} \gaterm + O(1),\\
&&K^{\rho}_{ab}=\frac{2(\xfu)^2 \overset{(0)}{h}_{ab} }{1+(\xfu)^2}+
\sqrt{\tau}\frac{2\xfd
    k^i_{ab}}{1+(\xfu)^2}+8\sqrt{\tau}\frac{\xfu \xsd
    +\frac{1}{2}\gaterm}{(1+(\xfu)^2)^2}-4\sqrt{\tau}\xfd k^i_{ab} +
O(\tau), \;\;\;\;\;\; \\
&&K^i_{\tau\tau}=-\frac{\xfu}{4\tau^{3/2}}-\frac{1}{2\tau}\frac{\xfu
    \xsd
    +\frac{1}{2}\gaterm}{1+(\xfu)^2}\xfu+\frac{1}{4\tau}
\overset{\text{\tiny{(0)}}}{\Gamma}{}^{i}_{km}\ 
\overset{\text{\tiny{(1)}}}{X}{}^k\overset{\text{\tiny{(1)}}}{X}{}^m +
O(\frac{1}{\tau^{1/2}}),\\
&&K^i_{ab}=-\frac{1}{\sqrt{\tau}}\frac{\overset{(0)}{h}_{ab}}{1+(\xfu)^2}
\xfu+k^i_{ab}-\frac{2}{1+(\xfu)^2}\overset{(0)}{h}_{ab}
\overset{\text{\tiny{(2)}}}{X}{}^i+\frac{\xfd k^i_{ab}
  \xfu}{1+(\xfu)^2} \nn\\
&& \qquad \qquad\qquad \qquad\qquad \qquad\qquad 
\qquad \qquad + 4 \frac{\xfu \xsd
  +\frac{1}{2}\gaterm}{(1+(\xfu)^2)^2}\xfu + O(\tau^{1/2}), 
\end{eqnarray}
Since $n^Q_\mu\frac{\partial X^{\mu}}{\partial\tau}=0$, we have
\begin{eqnarray}\label{normalrelation}
n^Q_\rho=-n^Q_i\frac{\partial X^i}{\partial \tau}.
\end{eqnarray}
The trace  $K$ is
\begin{eqnarray}\label{K}
K=-(p+1)\sqrt{\tau}n^Q_i\xfu-2\tau\frac{(p+1)(\xfu)^2+p}{1+(\xfu)^2}n^Q_j 
\overset{\text{\tiny{(2)}}}{X}{}^j+\tau n^Q_i k^i
+\tau\frac{n^Q_i\overset{\text{\tiny{(0)}}}{\Gamma}{}^{i}_{km}\ 
\overset{\text{\tiny{(1)}}}{X}{}^k\overset{\text{\tiny{(1)}}}{X}{}^m}{1+(\xfu)^2}
+ \cdots.
\end{eqnarray}

 Generally $n_\rho$ and $n_i$ can be expanded as
 \begin{eqnarray}\label{nE}
 n^Q_\rho=\frac{1}{\sqrt{\tau}}\overset{\text{\tiny{(0)}}}{n}{}^Q_\rho
+\overset{\text{\tiny{(1)}}}{n}{}^Q_\rho+ \cdots, \quad
n^Q_i=\frac{1}{\sqrt{\tau}}
\overset{\text{\tiny{(0)}}}{n}{}^Q_i+\overset{\text{\tiny{(1)}}}{n}{}^{Q}_i+
\cdots.
 \end{eqnarray}
Taking them to (\ref{normalrelation}) we have
\begin{eqnarray}
\overset{\text{\tiny{(0)}}}{n}{}^Q_\rho=-\frac{1}{2}
\overset{\text{\tiny{(0)}}}{n}{}^Q_i \xfu,\\
\overset{\text{\tiny{(1)}}}{n}{}^Q_\rho=-\frac{1}{2}
\overset{\text{\tiny{(1)}}}{n}{}^Q_i\xfu-\overset{\text{\tiny{(0)}}}{n}{}^Q_i
\overset{\text{\tiny{(2)}}}{X}{}^i.
\end{eqnarray}
Using the relation
\begin{eqnarray}
h_{\alpha \beta}=\frac{\partial X^\mu}{\partial x^{\alpha}}\frac{\partial
  X^\nu}{\partial x^{\beta}}\tilde{h}_{\mu\nu},
\end{eqnarray}
where 
$ds_Q^2 = h_{\a\b} d \t^\a d \t^\b = \tilde{h}_{\m\n} d X^\m dX^\n$ is
the induced metric on $Q$, we obtain that also $\tilde{h}_{\r i}=0$.
Hence
\begin{eqnarray}
h^{\tau\tau}=\tilde{h}^{\rho\rho}=G^{\rho\rho}-(n^Q_\rho)^2 G^{\rho\rho}
\end{eqnarray}
and as a result
\begin{eqnarray}\label{n1}
&& \overset{\text{\tiny{(0)}}}{n}{}^Q_i \xfu
=-\frac{\sqrt{(\xfu)^2}}{\sqrt{1+(\xfu)^2}},\\
&& \overset{\text{\tiny{(1)}}}{n}{}^Q_i \xfu=-2
\overset{\text{\tiny{(0)}}}{n}{}^Q_i
\overset{\text{\tiny{(2)}}}{X}{}^i-2\frac{\xfu \xsd
  +\frac{1}{2}\gaterm}{\sqrt{(\xfu)^2}(1+(\xfu)^2)^{3/2}}.
\end{eqnarray}
The gauge $h_{a\tau}=0$ lead to the result 
$\partial_a \overset{\text{\tiny{(0)}}}{X}_i\xfu=0$, which means
$\xfu$ is orthogonal to boundary submanifold $P$. Using
$n^Q_\mu\partial_a X^\mu=0$ one could show that 
$\overset{\text{\tiny{(0)}}}{n}{}^Q_i$ is also orthogonal to $P$. We
have the following relations
\begin{eqnarray}\label{np}
&&\xfu=\sqrt{(\xfu)^2} n^i,\\ \nonumber
&& \overset{\text{\tiny{(0)}}}{n}{}^Q_i=-\frac{1}{\sqrt{1+(\xfu)^2}}n_{i},
\end{eqnarray}
where $n_i$ is the unit normal vector of $P$.
Taking (\ref{nE})(\ref{n1})(\ref{np}) into (\ref{K}) we have
\begin{eqnarray}
K=-(p+1)\overset{\text{\tiny{(0)}}}{n}{}^Q_i \xfu+\sqrt{\tau} 
\overset{\text{\tiny{(0)}}}{n}{}^Q_i k^i-\sqrt{\tau }
\frac{2p}{1+(\xfu)^2}\overset{\text{\tiny{(0)}}}{n}{}^Q_i
\overset{\text{\tiny{(2)}}}{X}{}^i+p
\sqrt{\tau} \frac{\gaterm}{(1+(\xfu)^2)^{3/2}\sqrt{(\xfu)^2}}
+ O(\tau).
\end{eqnarray}
Now on the surface $Q$, $K=\frac{p+1}{p}T$, we obtain
\begin{eqnarray}
  &&n_i\xfu=\sqrt{\frac{T^2}{p^2-T^2}},\label{appax1}\\
  &&n_{i}\overset{\text{\tiny{(2)}}}{X}{}^i=\frac{1+(\xfu)^2}{2p}
  n_ik^i-\frac{1}{2} n_i\overset{\text{\tiny{(0)}}}{\Gamma}{}^{i}_{km}
  \overset{\text{\tiny{(1)}}}{X}{}^{k}\overset{\text{\tiny{(1)}}}{X}{}^{m}.
  \label{appax2}
\end{eqnarray}
Recall that from the gauge $h_{\tau a}=0$, we can solve the transverse
components of $X^i$ as eqs.(\ref{hat1},\ref{hat2}).
Combining eqs.(\ref{appax1},\ref{appax2}) and eqs.(\ref{hat1},\ref{hat2})
together, we recover
exactly
eqs.(\ref{finhat1},\ref{finhat2}).

\section{Boundary Weyl Anomaly for the Proposal of \cite{Takayanagi:2011zk}}

In this appendix, we show that the BC (\ref{NBC1}) proposed by
\cite{Takayanagi:2011zk} always make vanish the central charges
$c_2$ and $b_1$ in the boundary Weyl anomaly (\ref{3dBWA},\ref{4dBWA})
for 3d and 4d BCFT. Since $b_1$ is expected to satisfy a c-like theorem and
describes the degree of freedom on the boundary, thus it is important
for $b_1$ to be non-zero.   
We emphasis that this holds for any energy-momentum tensor
$T^Q_{\a\b}$ on $Q$ as long as the BC \eq{NBC1} holds. 
In this sense, the proposal of \cite{Takayanagi:2011zk} is too restrictive to
include the general BCFT, in particular, the non-trivial 4d BCFT.

Let us first start with a simple example 
to see explicitly how $c_2$ and $b_1$ vanish 
in the proposal of \cite{Takayanagi:2011zk}.  Consider $AdS$ with
cylindrical coordinates on $M$
eqs.(\ref{pureAdS4metric},\ref{pureAdS5metric3}) so that only the
$\text{Tr}\bar{k}^{d-1}$ terms are non-vanishing in the Weyl anomaly
(\ref{3dBWA},\ref{4dBWA}). 
We note that in the present case, the
equation \eq{NBC1} does not admit a solution with a constant $T$ term
and one needs to include on $Q$ either nontrivial matter fields  or higher
derivatives gravitational action terms. For simplicity, let us
consider the addition of an intrinsic Ricci
scalar $ R_Q$  on $Q$. In other words,
we focus on the action (\ref{generalaction}). Requiring all the
components of stress tensors on $Q$ vanishing, we get the following
exact solutions
\begin{eqnarray}\label{appsolution} 
\lambda=\frac{1}{2(d-2)} \coth \rho_*,\ \  T=(d-1) \coth (2 \rho_* ),
\ \ 
r=r_0+  \sinh\rho_*\ z  .
\end{eqnarray}
Substituting
eqs.(\ref{pureAdS4metric},\ref{pureAdS5metric3},\ref{appsolution})
into the action (\ref{generalaction}) and selecting the logarithmic
divergent term after integration alone $r$ and $z$, we find 
\begin{eqnarray}\label{appTii} 
\left<T^a_a\right>_P=0
\end{eqnarray}
for both 3d and 4d BCFT. This means that $c_2=b_1=0$. 
This example can be easily generalized to
include general higher curvature terms, i.e., we replace $R_Q$ by
$L(R_{Q \alpha\beta\lambda\gamma})$ in action
(\ref{generalaction}). Using the trick of \cite{Miao:2013nfa}, we
expand $L(R_{Q \alpha\beta\lambda\gamma})$ around a
`background-curvature' $\bar{R}_{Q
  \alpha\beta\lambda\gamma}=-\frac{1}{\cosh^2
  \rho_*}(h_{\alpha\lambda}h_{\beta\gamma}-h_{\alpha\gamma}h_{\beta\lambda})$. Then
we find only the first a few terms up to $(R_Q-\bar{R}_Q)^{d-1}\sim
z^{d-1}$ contribute to the boundary Weyl anomaly for $d$-dimensional
BCFT. We have worked out the cases for 3d and 4d BCFT on cylinders and
find they all yield eq.(\ref{appTii}). So the boundary Weyl anomaly
$c_2,b_1$ indeed vanish for 3d and 4d BCFT in the proposal of
\cite{Takayanagi:2011zk}. 
We have also
constructed a model with only matter on $Q$ (non-minimally coupled
scalar field with suitable potential energy), which also yield
$c_2=b_1=0$.

Now let us present the general proof. Consider the following action
\begin{eqnarray}\label{actionapp1}
I=\int_N \sqrt{G} (R-2 \Lambda) + 2\int_Q \sqrt{h} \left(
K-T+L_m(\phi) \right)+ 2\int_M \sqrt{g}\ K+2\int_P
\sqrt{\sigma}\ \theta
\end{eqnarray}
where $L_m(\phi)$ is the Lagrangian of matter fields $\phi$ on
$Q$. According to \cite{Henningson:1998gx}, we can derive the Weyl
anomaly as the logarithmic divergent term of the gravitational
action. Recall that $I_M$ and $I_P$ do not contribute the  logarithmic
divergent term \footnote{Instead of $\ln z$, $I_M$ and $I_P$ may
  contribute terms such as $z^n \ln z$ with $n>1$, which vanish in the
  limit $z \to 0$.}.
Considering the variation of the on-shell action, we have
\begin{eqnarray}\label{dactionapp1-0}
\delta I&=&-\int_Q \sqrt{h} \left(
(K^{\alpha\beta}-(K-T)h^{\alpha\beta}-\frac{1}{2} T^Q_{\alpha\beta})
\delta h_{\alpha\beta}+ E_{\phi} \delta \phi \right)
\nonumber\\ 
&&  \qquad\qquad -\int_M \sqrt{g} (K^{ij}-K h^{ij})\delta g_{ij}+\int_P
\sqrt{\sigma}\left( \theta \sigma^{ab}\delta \sigma_{ab} +
P_{\phi}\delta \phi \right),
\end{eqnarray}
where $E_{\phi}$ denotes E.O.M for matter fields $\phi$ on $Q$, 
$ P_{\phi}$ is the conjugate momentum of $\phi$ along the direction
$n_P$, which is the normal vector pointing from $Q$ to $P$. 
If one impose the BC (\ref{NBC1}), one obtain for
arbitrary boundary variations $\delta g_{ij}$, $\delta \sigma_{ab}$
and $\d \phi$:
\begin{eqnarray}\label{dactionapp1}
\delta I=-\int_M \sqrt{g} (K^{ij}-K h^{ij})\delta g_{ij}+\int_P
\sqrt{\sigma}\left( \theta \sigma^{ab}\delta \sigma_{ab} +
P_{\phi}\delta \phi \right)
\end{eqnarray}
where we have used the EOM  $E_{\phi}=0$.
It is worth noting that the integral on $Q$ vanishes due to the BC
(\ref{NBC1}). This is the main reason why the proposal of
\cite{Takayanagi:2011zk} yields trivial boundary central charges $c_2$
and $b_1$ in eqs.(\ref{3dBWA},\ref{4dBWA}). In fact as we will show below, the
integration on $M$ and $P$ in eq.(\ref{dactionapp1}) are not
sufficient to produce the full structures of the boundary Weyl anomaly. 

To proceed, we note that the logarithmic divergent term of $\delta I$ 
is 
equal to the variation of the Weyl anomaly $\mathcal{A}$
\begin{eqnarray}\label{dWeylapp1}
\delta I|_{\ln \epsilon}=\delta\mathcal{A}=\delta \int_M 
\sqrt{g_0}\left<T^i_i\right>_M +\delta \int_P \sqrt{\sigma_0} \left<T^a_a\right>_P.
\end{eqnarray}
Since there is no integration alone $z$ on $M$ and $P$, the only way
to produce $\ln z$ in $\delta I$ is that the integral element includes
$\ln z$. There are two possible sources for $\ln z$: one is the
expansion of $g_{ij}$ and the other one is the expansion of the
embedding function (\ref{Qsuface})
\begin{eqnarray}\label{twolnz1}
&&g_{ij}=g^{(0)}_{ij}+z^2 g^{(1)}_{ij}+\cdots + z^d( g^{(d/2)}_{ij}+
  h^{(d/2)}_{ij} \ln z)+ \cdots ,\quad \mbox{for even $d$} \\
&&x=a_1 z+ a_2 z^2+ \cdots + ( b_{d+1} \ln z+ a_{d+1}) z^{d+1}+ \cdots \label{twolnz2}.
\end{eqnarray}
Note that there is no $\ln z$ term in $g_{ij}$ when $d$ is odd. 
As a  result, there is no bulk Weyl anomaly $\left<T^i_i\right>_M$ for
odd $d$. It is also worth keeping in mind that $h^{(d/2)}_{ij} $ and $
b_{d+1}$ are of the same order $O(k^d)$ 
where $k$ is the trace of the extrinsic curvature of $P$.  
In general, E.O.M for matter
fields $E_{\phi}=0$ will also give $\ln z$ terms in $\phi$ . However,
such terms are expected to yield new contributions to Weyl anomaly in
addition to the geometric Weyl invariant such as
eqs.(\ref{3dBWA},\ref{4dBWA}). See \cite{deHaro:2000vlm,Nojiri:1999mh}
for some examples. Since here we are interested only in the geometric Weyl
invariant which defines $c_2$ and $b_1$, we will ignore these $\ln z$ terms
of $\phi$ (from $E_{\phi}=0$) in this appendix. Of course, $\phi$ can
inherit $\ln z$ terms from $g_{ij}$ (\ref{twolnz1}) and $x$
(\ref{twolnz2}) through eq.(\ref{NBC1}). And these  $\ln z$ terms are
functions of $h^{(d/2)}_{ij} $ and $ b_{d+1}$.

Let us firstly consider the case without the boundary $P$, i.e., the 
standard case of AdS/CFT. From the above discussions, we must have
\begin{eqnarray}\label{dIM} 
(\delta I)_M|_{\ln \epsilon}=-\int_M \sqrt{g} (K^{ij}-K h^{ij})\delta
  g_{ij}=\delta \int_M \sqrt{g_0}\left<T^i_i\right>_M 
\end{eqnarray}
 When $d$ is odd, we have $\left<T^i_i\right>_M=0$, which agrees with
 the fact that there is no $\ln z$ term in $g_{ij}$ and thus in $
 (\delta I)_M$. When $d$ is even, one can check eq.(\ref{dIM}) by
 straightforward calculations. Actually eq.(\ref{dIM}) must be
 satisfied since $I|_{\ln \epsilon}=\int_M
 \sqrt{g_0}\left<T^i_i\right>$ in AdS/CFT.

In the presence boundary $P$, the formulas of $(\delta I)_M$ and
$\delta( \sqrt{g_0}\left<T^i_i\right>_M )$ do not have any change. So
eq.(\ref{dIM}) is still satisfied up to a possible boundary term
$\Omega_M$ on $P$ from $\delta( \sqrt{g_0}\left<T^i_i\right>_M
)$. Then from eqs.(\ref{dactionapp1},\ref{dWeylapp1}), we get
\begin{eqnarray}\label{dIP} 
(\delta I)_P|_{\ln \epsilon}=\int_P \sqrt{\sigma}\left( \theta
  \sigma^{ab}\delta \sigma_{ab} + P_{\phi}\delta \phi \right)|_{\ln
    \epsilon}=\delta  \int_P \sqrt{\sigma_0} \left<T^a_a\right>_P +
  \int_P \sqrt{\sigma_0}\Omega_M(\delta g_{ij})
\end{eqnarray}
Notice that only the terms linear in $h^{(d/2)}_{ij} $ and $ b_{d+1}$
could include $\ln z$ in $(\delta I)_P$. However,  $h^{(d/2)}_{ij} $
and $ b_{d+1}$ are of order $O(k^d)$, while $\left<T^a_a\right>_P$ is
of order $O(k^{d-1})$. Thus they cannot contribute to
$\left<T^a_a\right>_P$ at all. Actually, the terms linear in
$h^{(d/2)}_{ij} $ and $ b_{d+1}$  take the form $z \ln z$, which
vanish in the limit $z\to 0$. Thus, we have
\begin{eqnarray}\label{keyresult} 
\delta  \int_P \sqrt{\sigma_0} \left<T^a_a\right>_P + \int_P
\sqrt{\sigma_0}\Omega_M(\delta g_{ij})=0
\end{eqnarray}
for
arbitrary boundary variations.

For 3d BCFT, $\left<T^i_i\right>_M =0 $ and $\Omega_M(\delta g_{ij})$
disappear. Eq.(\ref{keyresult}) implies that  $\int_P \sqrt{\sigma_0}
\left<T^a_a\right>_P$ is a topological invariant. As a result, we must
have $c_2=0$ in the boundary Weyl anomaly (\ref{3dBWA}). 
For 4d BCFT,  $\left<T^i_i\right>_M$  and $\Omega_M(\delta g_{ij})$ 
are non-zero. Note that $\Omega_M(\delta g_{ij})$  is proportional to
the Weyl tensor $C$ and its derivatives. Therefore for the simple
case where $C_{ijkl}|_P=\nabla_m C_{ijkl}|_P=0$, we have
$\Omega_M(\delta g_{ij})=0$. This together with 
eq.(\ref{keyresult})  implies that 
$\int_P \sqrt{\sigma_0} \left<T^a_a\right>_P$ is a topological
invariant. So $b_1$ related to $\text{Tr}\bar{k}^3$ must vanish in the
boundary Weyl anomaly (\ref{4dBWA}). Notice that in this argument 
we only require $C_{ijkl}$ to vanish at the boundary $P$. It can be
nontrivial inside $M$. For
instance, the following metric $g^{(0)}_{ij}$ with a free parameter
$c$ works well for our purpose:
\begin{eqnarray}\label{ds0app} 
ds_0^2=dx^2+[(1+c x)^2 +O(x^4)]dy_1^2+ [1+O(x^4)]dy_a^2,
\end{eqnarray}
where $x=0$ denotes the location of $P$. 
One can easily check that the above metric satisfy
$C_{ijkl}|_P=\nabla_m C_{ijkl}|_P=0$ but $C_{ijkl}|_M\ne 0$ generally.
Now since the boundary central charges are
independent of the shapes of the boundary, so we also have $b_1=0$ for
the boundary with $C_{ijkl}|_P\ne 0$. 
One can also include higher curvature terms  on $Q$ in the action
\eq{actionapp1} and the proof proceeds
exactly the same way. 
Therefore we find that, independent of the form of the matter or
gravitational action,  the proposal of \cite{Takayanagi:2011zk}
always give $c_2=b_1=0$ in the boundary Weyl anomaly
(\ref{3dBWA},\ref{4dBWA}).
As we explained above, the reason why the
proposal of \cite{Takayanagi:2011zk} always yield $c_2=b_1=0$ is that
the requirement that all the components of stress tensors on $Q$
vanish $T_{\alpha\beta}=0$ is too strong. On the other hand, 
if one require only the trace of the stress tensor to vanish as in our proposal
then the integral on $Q$ 
in eq.(\ref{dactionapp1-0})
 is no longer
zero and
one can indeed obtain non-trivial boundary central charges $c_2$ and $b_1$ in
eqs.(\ref{3dBWA},\ref{4dBWA}).

Finally we remark that, as the 
careful readers may notice also,  the solution (\ref{appsolution}) with
$a_2=0$ 
does not obey the universal law for
$a_2=-\frac{\cosh^2\rho_*}{2(d-1)}\text{Tr}k$ as in eq.(\ref{uniTa}).  
This is not surprising since the parameter $\lambda$ does not lie
in the ``physical range''. In fact the solutions to our
proposal $T_{BY}{}^{\alpha}_{\alpha}=0$ are not unique when we allow
higher curvature terms in the stress tensors. 
Generally as long as the parameters of the
higher curvature terms lie in some ``physical'' region,
there is an unique solution which 
satisfies
the universal law for $a_2$ and give
the non-trivial boundary central charges. We select this kind
of solution as the physical one. However when one
set the parameters of higher curvature terms to the critical value as
in eq.(\ref{appsolution}), the physical solution is replaced by a
different solution which violate the
universal law of
$a_2=-\frac{\cosh^2\rho_*}{2(d-1)}\text{Tr}k$. 
Actually, the same situation already
appears in \cite{Schwimmer:2008yh}: for higher curvature gravity such
as Lovelock gravity, the bulk entangling surfaces obtained by
minimizing the entropy functional are not unique. One usually select
the one which can be continuously reduced to the minimal surface
when the parameters of higher curvature terms are all turned off. This
kind of surface always satisfy 
an universal relation for $a_2$
\cite{Schwimmer:2008yh}. However if one
set the parameters of higher curvature terms to the critical value as
in eq.(\ref{appsolution}), there exist solutions which violate the
universal relation
\cite{Schwimmer:2008yh}. Thus, the universality of $a_2$ in our
proposal has the same meaning as the one in \cite{Schwimmer:2008yh}:
it  holds as long as 
the parameters of  higher curvature terms 
lie in the physical
ranges. 
Curiously, the proposal of \cite{Takayanagi:2011zk} has
solution only if the parameter of higher curvature terms takes the
critical value $\lambda=\frac{1}{2(d-2)} \coth \rho_*$ and this 
prevents the realization of non-trivial boundary central charges.

\section{Derivations of Boundary Contributions to Weyl Anomaly}

In sect.2.2, we have shown the key steps of  holographic derivations
of boundary contributions to Weyl anomaly. Here we provide more
details. We work in Gaussian normal coordinate and find the following
formulas useful:
\begin{eqnarray}\label{appCformulas1} 
&& T=(d-1) \tanh \rho_*,\nonumber\\
  && x= \sinh\rho_*\; z-\frac{k\cosh^2\rho_* }{2(d-1)} \;
  z^2+ a_3 z^3+a_4 z^4+\cdots .
\end{eqnarray}
Since we want to consider the general boundary condition
(\ref{mixedBCN1}), we
keep $a_3$ and $a_4$ off-shell in this appendix.
For the bulk action $I_N=-2d \int_N \sqrt{G}$ and the BCFT boundary
metric \eq{BCFTmetric}, we
have
\be
\sqrt{G}=\frac{1}{z^{d+1}}\sqrt{ \overset{(0)}{g}}
\Big[ 1+ \frac{1}{2} z^2 \overset{(1)}{g}{}^i_i+\cdots \Big],
\ee
where
\begin{eqnarray}\label{appCformulas2} 
 && \sqrt{ \overset{(0)}{g}} =\sqrt{ \overset{(0)}{\sigma}}
      \Big[ 1+ k x+\frac{1}{2}  \left(k^2+q-2
        \text{Tr}k^2\right)x^2 \nn \\
        &&  \qquad \qquad  \qquad \qquad +\frac{1}{6}
        \left(k^3+3 k q-6 k\text{Tr}k^2+3 l+8 \text{Tr}k^3-6
        \text{Tr}(kq)\right)x^3+
        \cdots \Big],\nonumber\\
 && \overset{(1)}{g}{}^i_i = -\frac{\mathcal{R}-k^2-2 q+3
        \text{Tr}k^2}{2 (d-1)}+
      \frac{-2 k \text{Tr}k^2+kq+3 l+6
        \text{Tr}k^3+\text{Tr}(k\mathcal{R})-5
        \text{Tr}(kq)}{d-1}x+\cdots .\;\;\;\;\;\;\;\;\;
\end{eqnarray}
Here $\cdots$ denotes
terms of order $O(k^4)$ which do not contribute
to boundary Weyl anomaly for 3d BCFT and 4d BCFT.

For the boundary action $I_Q=2\int_Q \sqrt{h}(K-T)$, we need
\begin{eqnarray}\label{appCformulas3} 
&&\sqrt{h}=\frac{1}{z^{d}}\sqrt{ \overset{(0)}{g}}\sqrt{1+ g_{xx}
    x'^2} [1+ \frac{1}{2}
    z^2 \overset{(1)}{g}{}^a_a ]+\cdots, \nonumber\\
&& g_{xx}=1+ \frac{\mathcal{R}+2 (d-2) q-2 d \text{Tr}k^2-k^2+5
    \text{Tr}k^2}{2 (d-2)
    (d-1)}z^2 \nonumber\\
  && \ \ \  \ \ \ -\frac{-3 d l-4 d \text{Tr}k^3+4 d \text{Tr}(kq)+k q
    -2 k \text{Tr}k^2+6 l+10 \text{Tr}k^3-9 \text{Tr}(kq)
    +\text{Tr}(k\mathcal{R})}{d^2-3 d+2} z^2 x+\cdots, \nonumber\\
  && g^{xx}=1- \frac{\mathcal{R}+2 (d-2) q-2 d \text{Tr}k^2-k^2+5
    \text{Tr}k^2}{2 (d-2) (d-1)}z^2\nonumber\\
  && \ \ \  \ \ \ +\frac{-3 d l-4 d \text{Tr}k^3+4 d \text{Tr}(kq)
    +k q-2 k \text{Tr}k^2+6 l+10 \text{Tr}k^3-9 \text{Tr}(kq)
    +\text{Tr}(k\mathcal{R})}{d^2-3 d+2} z^2 x+\cdots, \nonumber\\
  && \overset{(1)}{g}{}^a_a=\frac{\mathcal{R}-k^2+\text{Tr}k^2}{4-2 d}
  +\frac{k q-2 k \text{Tr}k^2+2 \text{Tr}k^3-\text{Tr}(kq)
    +\text{Tr}(k\mathcal{R})}{d-2}x
  +\cdots \nonumber\\
    && K=\frac{1}{\sqrt{h}}\partial_{\mu}( \sqrt{h} n^u ), \ \ \ \
    n^u= \frac{z}{\sqrt{g^{xx}+x'^2}} (-x', g^{xx},
    \overset{(0)}{\sigma}{}^{ab}\partial_b x )+\cdots, 
\end{eqnarray}
where $x'=\partial_z x$ and
$\cdots$ denotes
higher order
terms irrelevant to the  boundary Weyl anomaly for 3d BCFT and 4d BCFT.

Now we are ready to derive the boundary Weyl anomaly.  Substituting
the above formulas into the action (\ref{action1}) and
selecting the logarithmic divergent terms after the integral along $x$
and $z$, we can obtain the boundary Weyl anomaly.  For 3d BCFT, we have
\begin{eqnarray}\label{appC3dBCFTINIQ} 
&&I_N=-\frac{1}{8} \ln\frac{1}{\epsilon} \int_P \sqrt{
    \overset{(0)}{\sigma}} \Big[48 a_3-3 \sinh\rho_* \left(2
    \mathcal{R}+k^2-2 q+2 \text{Tr}k^2\right)+\sinh (3 \rho_* )
    \left(-k^2+2 q-4 \text{Tr}k^2\right)\Big],\ \ \ \ \ \nonumber\\
&&I_Q=\frac{1}{8} \ln\frac{1}{\epsilon} \int_P \sqrt{
    \overset{(0)}{\sigma}}\Big[48 a_3+\sinh (\rho ) \left(2
    \mathcal{R}+k^2+6 q-14 \text{Tr}k^2\right)+\sinh (3 \rho_* )
    \left(-k^2+2 q-4 \text{Tr}k^2\right)\Big], \nonumber
\end{eqnarray}
where we have ignored terms without $\ln\frac{1}{\epsilon}$
above. Combining $I_N$ and $I_Q$ together, we get
\begin{eqnarray}\label{appC3dBCFTI} 
  I|_{\ln\frac{1}{\epsilon}}= \int_P \sqrt{ \overset{(0)}{\sigma}}
  (\mathcal{R}- \text{Tr}\bar{k}^2 )\sinh\rho_*,
\end{eqnarray}
which exactly gives the boundary Weyl anomaly (\ref{3dBWA1}).
It is remarkable that $a_3$ and all non-conformal invariant terms
automatically cancel each other out.

Similarly, for 4d BCFT we have
\begin{eqnarray}\label{appC4dBCFTIN} 
  &&I_N= \ln\frac{1}{\epsilon} \int_P\sqrt{ \overset{(0)}{\sigma}}
  \frac{1}{72}\Big[-576 a_3 k \sinh\rho_*-576 a_4-2 k^3+23 k q-16 k
    \mathcal{R} \nonumber\\
    &&\ \ \  \ \ \ \ \ \ \ \ \ \ \ \  \ \ \ \ \ \ \ \ \ \ \ \ \
    -30 k \text{Tr}k^2+45 l+72 \text{Tr}k^3-66
    \text{Tr}(kq)+24
    \text{Tr}(k\mathcal{R}) \nonumber\\
    &&\ \ \  \ \ \ \ \ \ \ \ \ \ \ \  \ \ \ \ \ \ \ \ \  \ \ \ \ \  \
    -4 \cosh (2 \rho_* ) (k q-2k \mathcal{R}+9
    l+ 12 \text{Tr}k^3-12 \text{Tr}(kq)+6\text{Tr}(k\mathcal{R}))
    \nonumber\\
&&\ \ \  \ \ \ \ \ \ \ \ \ \ \ \  \ \ \ \ \ \ \ \ \  \ \ \ \ \  \ +
    \cosh (4 \rho_* ) \left(2 k^3-3 k q+6 k \text{Tr}k^2-9 l-24
    \text{Tr}k^3+18 \text{Tr}(kq)\right)
    \Big] \label{appC4dBCFTIQ} \\
&&I_Q=\ln\frac{1}{\epsilon} \int_P\sqrt{ \overset{(0)}{\sigma}}
  \frac{1}{216}\Big{[}1728 a_3 k \sinh\rho_*+1728a_4+26 k^3-135 l-69 k
    q-60 k \mathcal{R}\nonumber\\
    &&\ \ \  \ \ \ \ \ \ \ \ \   \ \ \ \ 
    +54 k \text{Tr}k^2-216 \text{Tr}k^3+198
    \text{Tr}(kq)+144
    \text{Tr}(k\mathcal{R}) \nonumber\\
    &&\ \ \  \ \ \ \ \ \ \ \ \ \ \ \ \
    +12 \cosh (2 \rho_*) \left(k^3+4
   k q+k\mathcal{R}-9 k\text{Tr}k^2+9 l+30 \text{Tr}k^3-21
    \text{Tr}(kq)-3 \text{Tr}(k\mathcal{R})\right) \nonumber\\    
&&\ \ \  \ \ \ \ \ \ \ \ \ \ \ \  \  +3
    \cosh (4 \rho_*) \left(-2k^3-6 k \text{Tr}k^2+24 \text{Tr}k^3-18
    \text{Tr}(kq)+3 k q+9 l\right) \Big{]}
\end{eqnarray}
Combining the above $I_N$ and $I_Q$ together, we obtain
\begin{eqnarray}\label{appC4dBCFTI} 
&&I|_{\ln\frac{1}{\epsilon}}=\int_P \sqrt{
    \overset{(0)}{\sigma}}\frac{1}{54}
  \Big{[}5 k^3-9 k (3 \mathcal{R}+\text{Tr}k^2)+54
    \text{Tr}(k\mathcal{R})\ \ \ \ \ \ \ \ \ \ \ \ \ \  \ \nonumber\\
    &&\ \ \  \ \ \ \ \ \ \ \ \  + 3 \cosh (2 \rho_* )
    \left(k^3+3 k (q+\mathcal{R}-3 \text{Tr}k^2)+18 \text{Tr}k^3-9
    \text{Tr}(kq)-9\text{Tr}(k\mathcal{R})\right)\Big{]}\nonumber\\
  &&= \int_P \sqrt{ \overset{(0)}{\sigma}} \
  \Big{[}\ \frac{1}{8} E_4^{\rm bdy}+ (\cosh(2\rho_*)-\frac{1}{3})
  \text{Tr} \bar{k}^3 -\cosh(2\rho_*) C^{ac}_{\ \ \ b c}
  \bar{k}_{\ a}^b \ \Big{]}
\end{eqnarray}
which
yields exactly the boundary Weyl anomaly (\ref{4dBWA1}). In the
above calculations, we have used
eqs.(\ref{foumulasbird1},\ref{foumulasbird2},\ref{foumulasbird3}). Similar
to the 3d case, $a_3$, $a_4$ and all of the non-conformal invariant
terms automatically cancel each other out in the final results.

To end this appendix, let us discuss the physical meaning of the
parameter $\rho_*$.  As we have mentioned, $\rho_*$ can be regarded as
the holographic dual of boundary conditions of BCFT since it affects
the boundary
entropy \cite{Takayanagi:2011zk} and also the boundary central charges
(\ref{3dBWA1},\ref{4dBWA1})
which are closely related to the boundary conditions of BCFT. To cover
the general boundary condition, it is natural to keep $\rho_*$ free
rather than to set it zero. If we set $\rho_*=0$, we get zero boundary
entropy $S_{bdy}=\frac{\rho_*}{4G_N}$ for 2d BCFT
\cite{Takayanagi:2011zk} which gives trivial BCFT. Furthermore, it is
expected that the boundary central charges related to different
conformal invariants  are independent in general. As a result we must
keep $\rho_*$ free. Of course, as discussed in sect.4  one could add
intrinsic curvature terms on $Q$ in order to make all the boundary
central charges independent.

Finally, we notice that for 4d BCFT, the case $\rho_*=0$ can
reproduce the proposal of \cite{Solodukhin:2015eca} and agree with the
boundary Weyl anomaly of $\mathcal{N}=4$ super Yang-Mills multiplet
with a special choice of boundary conditions that preserve half of
supersymmetry \cite{Astaneh:2017ghi}. For the convenience of the
reader, we list the boundary Weyl anomaly of free super Yang-Mills
multiplet with general boundary condition in the large $N$ limit below
\cite{Astaneh:2017ghi}.
\begin{eqnarray}\label{freesuperYangMills}
\left<T^a_a\right>_P= \frac{1}{8} E_4^{\rm bdy}+ (\frac{2}{3}+\frac{\Delta n}{70})
  \text{Tr} \bar{k}^3 -C^{ac}_{\ \ \ b c}
  \bar{k}_{\ a}^b,
\end{eqnarray}
 where $\Delta n :=n_s^D-n_s^R$ with the total number $n_s^D+n_s^R=6$
 fixed.
 Here `s' denotes scalar, `D' and `R'  refers to the Dirichlet boundary
 condition and Robin boundary condition respectively. $\Delta n=0$
 corresponds to the case that  half of the supersymmetry is preserved
 \cite{Gaiotto:2008sa,Gaiotto:2008ak,Hashimoto:2014nwa}. 
It is not known in general when  non-renormalization theorem of the
trace anomaly holds. In case it does, 
the result (\ref{freesuperYangMills})  agrees
 with the general expression \eq{4dBWA1genRRR} of the
 holographic anomaly if the coefficients for the intrinsic curvature terms on
 $Q$ are fixed to be:
\begin{eqnarray}\label{holographicandfreeCFT}
\lambda&=& \frac{1}{4} \tanh (2 \rho_* ),\\
\lambda_2&=& -\frac{\cosh ^2\rho_*  \coth ^3\rho_* }{1120}\Delta n ,
\end{eqnarray}
where $\rho_*$ is a free parameter.


\begin{thebibliography}{00}


\bibitem{Cardy:2004hm}
  J.~L.~Cardy,
  hep-th/0411189.

\bibitem{Maldacena:1997re}
  J.~M.~Maldacena,
  Int.\ J.\ Theor.\ Phys.\  {\bf 38}, 1113 (1999)
  [Adv.\ Theor.\ Math.\ Phys.\  {\bf 2}, 231 (1998)]
  [hep-th/9711200].

\bibitem{Takayanagi:2011zk}
  T.~Takayanagi,
  Phys.\ Rev.\ Lett.\  {\bf 107} (2011) 101602
  [arXiv:1105.5165 [hep-th]].

\bibitem{uvir}
A.~W.~Peet and J.~Polchinski,
  Phys.\ Rev.\ D {\bf 59} (1999) 065011
  doi:10.1103/PhysRevD.59.065011
  [hep-th/9809022].

\bibitem{Nozaki:2012qd}
  M.~Nozaki, T.~Takayanagi and T.~Ugajin,
  JHEP {\bf 1206} (2012) 066
  [arXiv:1205.1573 [hep-th]].

\bibitem{Fujita:2011fp}
  M.~Fujita, T.~Takayanagi and E.~Tonni,
  JHEP {\bf 1111} (2011) 043
  [arXiv:1108.5152 [hep-th]].


\bibitem{Jensen:2015swa} 
  K.~Jensen and A.~O'Bannon,
  Phys.\ Rev.\ Lett.\  {\bf 116}, no. 9, 091601 (2016)
  doi:10.1103/PhysRevLett.116.091601
  [arXiv:1509.02160 [hep-th]].

\bibitem{Erdmenger:2015spo} 
  J.~Erdmenger, M.~Flory, C.~Hoyos, M.~N.~Newrzella and J.~M.~S.~Wu,
  Fortsch.\ Phys.\  {\bf 64}, 109 (2016)
  doi:10.1002/prop.201500099
  [arXiv:1511.03666 [hep-th]].

\bibitem{Erdmenger:2014xya} 
  J.~Erdmenger, M.~Flory and M.~N.~Newrzella,
  JHEP {\bf 1501}, 058 (2015)
  doi:10.1007/JHEP01(2015)058
  [arXiv:1410.7811 [hep-th]].

\bibitem{Miyaji:2014mca} 
  M.~Miyaji, S.~Ryu, T.~Takayanagi and X.~Wen,
  JHEP {\bf 1505}, 152 (2015)
  doi:10.1007/JHEP05(2015)152
  [arXiv:1412.6226 [hep-th]].

\bibitem{Jensen:2013lxa} 
  K.~Jensen and A.~O'Bannon,
  Phys.\ Rev.\ D {\bf 88}, no. 10, 106006 (2013)
  doi:10.1103/PhysRevD.88.106006
  [arXiv:1309.4523 [hep-th]].

\bibitem{Estes:2014hka} 
  J.~Estes, K.~Jensen, A.~O'Bannon, E.~Tsatis and T.~Wrase,
  JHEP {\bf 1405}, 084 (2014)
  doi:10.1007/JHEP05(2014)084
  [arXiv:1403.6475 [hep-th]].

\bibitem{Gaiotto:2014gha} 
  D.~Gaiotto,
  arXiv:1403.8052 [hep-th].


\bibitem{Fursaev:2016inw} 
  D.~V.~Fursaev and S.~N.~Solodukhin,
  Phys.\ Rev.\ D {\bf 93}, no. 8, 084021 (2016)
  doi:10.1103/PhysRevD.93.084021
  [arXiv:1601.06418 [hep-th]].

\bibitem{Berthiere:2016ott} 
  C.~Berthiere and S.~N.~Solodukhin,
  Nucl.\ Phys.\ B {\bf 910}, 823 (2016)
  doi:10.1016/j.nuclphysb.2016.07.029
  [arXiv:1604.07571 [hep-th]].

\bibitem{He:2014gva} 
  S.~He, T.~Numasawa, T.~Takayanagi and K.~Watanabe,
  JHEP {\bf 1505}, 106 (2015)
  doi:10.1007/JHEP05(2015)106
  [arXiv:1412.5606 [hep-th]].


\bibitem{Hayward:1993my}
  G.~Hayward,
  Phys.\ Rev.\ D {\bf 47} (1993) 3275.

\bibitem{Miao:2017gyt} 
  R.~X.~Miao, C.~S.~Chu and W.~Z.~Guo,
  arXiv:1701.04275 [hep-th].


\bibitem{Rafael}
 López,~Rafael,
  ``Constant Mean Curvature Surfaces with Boundary,''
  Phys.\ Rev.\ D {\bf 47} (1993) 3275.


\bibitem{Herzog:2015ioa}
  C.~P.~Herzog, K.~W.~Huang and K.~Jensen,
  JHEP {\bf 1601}, 162 (2016)
  [arXiv:1510.00021 [hep-th]].

\bibitem{Fursaev:2015wpa}
  D.~Fursaev,
  JHEP {\bf 1512}, 112 (2015)
  [arXiv:1510.01427 [hep-th]].

\bibitem{Solodukhin:2015eca}
  S.~N.~Solodukhin,
  Phys.\ Lett.\ B {\bf 752}, 131 (2016)
  [arXiv:1510.04566 [hep-th]].

\bibitem{Graham:1999pm} 
  C.~R.~Graham and E.~Witten,
  Nucl.\ Phys.\ B {\bf 546}, 52 (1999)
  doi:10.1016/S0550-3213(99)00055-3
  [hep-th/9901021].


\bibitem{Schwimmer:2008yh}
  A.~Schwimmer and S.~Theisen,
  Nucl.\ Phys.\ B {\bf 801} (2008) 1
  [arXiv:0802.1017 [hep-th]].



\bibitem{Henningson:1998gx}
  M.~Henningson and K.~Skenderis,
  JHEP {\bf 9807} (1998) 023
  [hep-th/9806087].

\bibitem{Imbimbo:1999bj}
  C.~Imbimbo, A.~Schwimmer, S.~Theisen and S.~Yankielowicz,
  Class.\ Quant.\ Grav.\  {\bf 17} (2000) 1129
  [hep-th/9910267].

\bibitem{Miao:2013nfa}
  R.~X.~Miao,
  Class.\ Quant.\ Grav.\  {\bf 31}, 065009 (2014)
  [arXiv:1309.0211 [hep-th]].

\bibitem{Miao:2015iba}
  R.~X.~Miao,
  JHEP {\bf 1510} (2015) 049
  [arXiv:1503.05538 [hep-th]].



\bibitem{Huang:2016rol}
  K.~W.~Huang,
  JHEP {\bf 1608}, 013 (2016)
  [arXiv:1604.02138 [hep-th]].

\bibitem{Dong:2016wcf}
  X.~Dong,
  Phys.\ Rev.\ Lett.\  {\bf 116} (2016) no.25,  251602
  [arXiv:1602.08493 [hep-th]].

\bibitem{Lee:2014zaa}
  J.~Lee, A.~Lewkowycz, E.~Perlmutter and B.~R.~Safdi,
  JHEP {\bf 1503}, 075 (2015)
  [arXiv:1407.7816 [hep-th]].

\bibitem{Hung:2014npa}
  L.~Y.~Hung, R.~C.~Myers and M.~Smolkin,
  JHEP {\bf 1410}, 178 (2014)
  doi:10.1007/JHEP10(2014)178
  [arXiv:1407.6429 [hep-th]].

\bibitem{Chu:2016tps}
  C.~S.~Chu and R.~X.~Miao,
  JHEP {\bf 1612}, 036 (2016)
  doi:10.1007/JHEP12(2016)036
  [arXiv:1608.00328 [hep-th]].

\bibitem{Balasubramanian:1999re}
  V.~Balasubramanian and P.~Kraus,
  Commun.\ Math.\ Phys.\  {\bf 208} (1999) 413
  [hep-th/9902121].


\bibitem{deHaro:2000vlm}
  S.~de Haro, S.~N.~Solodukhin and K.~Skenderis,
  Commun.\ Math.\ Phys.\  {\bf 217} (2001) 595
  [hep-th/0002230].



\bibitem{Ryu:2006bv}
  S.~Ryu and T.~Takayanagi,
  Phys.\ Rev.\ Lett.\  {\bf 96} (2006) 181602
  [hep-th/0603001].


\bibitem{Lewkowycz:2013nqa}
  A.~Lewkowycz and J.~Maldacena,
  JHEP {\bf 1308} (2013) 090
  [arXiv:1304.4926 [hep-th]].

\bibitem{Jafferis:2015del}
  D.~L.~Jafferis, A.~Lewkowycz, J.~Maldacena and S.~J.~Suh,
  JHEP {\bf 1606}, 004 (2016)
  [arXiv:1512.06431 [hep-th]].

\bibitem{Dong:2016eik}
  X.~Dong, D.~Harlow and A.~C.~Wall,
  Phys.\ Rev.\ Lett.\  {\bf 117}, no. 2, 021601 (2016)
  [arXiv:1601.05416 [hep-th]].

\bibitem{McAvity:1993ue}
  D.~M.~McAvity and H.~Osborn,
  Nucl.\ Phys.\ B {\bf 406}, 655 (1993)
  [hep-th/9302068].

\bibitem{Song:2016pwx} 
  W.~Song, Q.~Wen and J.~Xu,
  Phys.\ Rev.\ Lett.\  {\bf 117}, no. 1, 011602 (2016)
  [arXiv:1601.02634 [hep-th]].

\bibitem{Herzog:2016kno}
  C.~Herzog and K.~W.~Huang,
  arXiv:1610.08970 [hep-th].

\bibitem{Hung:2011nu} 
  L.~Y.~Hung, R.~C.~Myers, M.~Smolkin and A.~Yale,
  JHEP {\bf 1112}, 047 (2011)
  doi:10.1007/JHEP12(2011)047
  [arXiv:1110.1084 [hep-th]].


\bibitem{Dong:2016fnf} 
  X.~Dong,
  Nature Commun.\  {\bf 7}, 12472 (2016)
  doi:10.1038/ncomms12472
  [arXiv:1601.06788 [hep-th]].

\bibitem{Donnelly:2014fua}
  W.~Donnelly and A.~C.~Wall,
  Phys.\ Rev.\ Lett.\  {\bf 114}, no. 11, 111603 (2015)
  [arXiv:1412.1895 [hep-th]].

\bibitem{Huang:2014pfa}
  K.~W.~Huang,
  Phys.\ Rev.\ D {\bf 92}, no. 2, 025010 (2015)
  doi:10.1103/PhysRevD.92.025010
  [arXiv:1412.2730 [hep-th]].

\bibitem{Bueno:2015rda}
  P.~Bueno, R.~C.~Myers and W.~Witczak-Krempa,
  Phys.\ Rev.\ Lett.\  {\bf 115}, 021602 (2015)
  [arXiv:1505.04804 [hep-th]].

\bibitem{Mezei:2014zla}
  M.~Mezei,
  Phys.\ Rev.\ D {\bf 91}, no. 4, 045038 (2015)
  [arXiv:1411.7011 [hep-th]].

\bibitem{Numasawa:2016emc}
  T.~Numasawa, N.~Shiba, T.~Takayanagi and K.~Watanabe,
  JHEP {\bf 1608}, 077 (2016)
  [arXiv:1604.01772 [hep-th]].

\bibitem{Nojiri:1999mh} 
  S.~Nojiri and S.~D.~Odintsov,
  Int.\ J.\ Mod.\ Phys.\ A {\bf 15}, 413 (2000)
  doi:10.1142/S0217751X00000197
  [hep-th/9903033].

\bibitem{Astaneh:2017ghi} 
  A.~F.~Astaneh and S.~N.~Solodukhin,
  arXiv:1702.00566 [hep-th].

\bibitem{Gaiotto:2008sa} 
  D.~Gaiotto and E.~Witten,
  J.\ Statist.\ Phys.\  {\bf 135}, 789 (2009)
  doi:10.1007/s10955-009-9687-3
  [arXiv:0804.2902 [hep-th]].

\bibitem{Gaiotto:2008ak} 
  D.~Gaiotto and E.~Witten,
  Adv.\ Theor.\ Math.\ Phys.\  {\bf 13}, no. 3, 721 (2009)
  doi:10.4310/ATMP.2009.v13.n3.a5
  [arXiv:0807.3720 [hep-th]].

\bibitem{Hashimoto:2014nwa} 
  A.~Hashimoto, P.~Ouyang and M.~Yamazaki,
  JHEP {\bf 1410}, 108 (2014)
  doi:10.1007/JHEP10(2014)108
  [arXiv:1406.5501 [hep-th], arXiv:1406.5501].





\end{thebibliography}
\end{document}